\documentclass{IEEEoj}
\def\BibTeX{{\rm B\kern-.05em{\sc i\kern-.025em b}\kern-.08em
    T\kern-.1667em\lower.7ex\hbox{E}\kern-.125emX}}
\AtBeginDocument{\definecolor{ojcolor}{cmyk}{0.93,0.59,0.15,0.02}}
\def\OJlogo{\vspace{-11pt}\includegraphics[height=20pt]{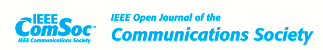}}

\usepackage{amsmath,amsthm,relsize}
\usepackage{amsfonts, amssymb, cuted}
\usepackage{cleveref}
\usepackage{mathtools}
\usepackage{algorithm}
\usepackage[noend]{algpseudocode}
\usepackage{cite}
\usepackage{xcolor}
\usepackage{soul}
\usepackage{graphicx}
\usepackage{tikz}
\usepackage{pgfplotstable}
\usepackage{pgfplots}
\usepackage{nicefrac}
\usepackage{support-caption}
\usepackage{subcaption}
\usepackage[font=footnotesize]{caption}
\usepackage{multirow}
\pgfplotsset{compat=1.15}
\usepackage{relsize}
\usetikzlibrary{patterns}
\usepackage{pifont}
\theoremstyle{definition}

\newcommand{\Bh}[0]{{\mathbf{h}}}

\newcommand{\Bv}[0]{{\mathbf{v}}}

\newcommand{\Bx}[0]{{\mathbf{x}}}

\newcommand{\rev}[1]{{\color{black} #1}}

\begin{document}

\receiveddate{XX Month, XXXX}
\reviseddate{XX Month, XXXX}
\accepteddate{XX Month, XXXX}
\publisheddate{XX Month, XXXX}
\currentdate{XX Month, XXXX}
\doiinfo{OJCOMS.2022.1234567}

\title{\huge{Enhancing Next-Generation Extended Reality Applications with Coded Caching}}
%\thanks{This work is supported by the Academy of Finland under grants no. 318927 (6Genesis Flagship), 319059 (%Coded Collaborative Caching for Wireless Energy Efficiency - 
%CCCWEE), and 
%343586 (%Cache-aided mmWave Access for Immersive Digital Environments - 
%CAMAIDE), and by the Finnish Research Impact Foundation %(Vaikuttavuuss\"a\"ati\"o) 
%under the project %Directional Data Delivery for Wireless Immersive Digital Environments (
%3D-WIDE%)
%.}}
\date{February 2023}

\author{M. SALEHI\authorrefmark{1}, K. HOOLI\authorrefmark{2}, J. HULKKONEN\authorrefmark{2}, AND A. T\"OLLI\authorrefmark{1}}
\affil{Center for Wireless Communications (CWC), P.O. Box 4500, FI-90014 University of Oulu, Finland}
\affil{Nokia Standards, 90620 Oulu, Finland}
\corresp{CORRESPONDING AUTHOR: M. Salehi (e-mail: mohammadjavad.salehi@oulu.fi).}
\authornote{This work was supported by the Academy of Finland under grants no. 318927 (6Genesis Flagship) and 343586 (CAMAIDE), and by the Finnish Research Impact Foundation under the project 3D-WIDE.}
\markboth{Enhancing Next-Generation Extended Reality Applications with Coded Caching}{Salehi, M. \textit{et al.}}

%\author{
%\IEEEauthorblockN{MohammadJavad~Salehi, Kari Hooli, Jari Hulkkonen, and Antti~T\"olli}
        
%\thanks{MohammadJavad Salehi and Antti T\"olli are with the Center for Wireless Communications (CWC), P.O. Box 4500, FI-90014 University of Oulu, Finland, e-mail address: \{first\_name.last\_name\}@oulu.fi. Kari Hooli and Jari Hulkkonen are with Nokia Standards, 90620 Oulu, Finland, e-mail address: \{first\_name.last\_name\}@nokia.com.}%
%}

\begin{abstract}
The next evolutionary step in human-computer interfaces will bring forward immersive digital experiences that submerge users in a 3D world while allowing them to interact with virtual or twin objects. Accordingly, various collaborative extended reality (XR) applications are expected to emerge, imposing stringent performance requirements on the underlying wireless connectivity infrastructure. In this paper, we examine how novel multi-antenna coded caching (CC) techniques can facilitate high-rate low-latency communications and improve users' quality of experience (QoE) in our envisioned multi-user XR scenario. Specifically, we discuss how these techniques make it possible to prioritize the content relevant to wireless bottleneck areas while enabling the cumulative cache memory of the users to be utilized as an additional communication resource. In this regard, we first explore recent advancements in multi-antenna CC that facilitate the efficient use of distributed in-device memory resources. Then, we review how XR application requirements are addressed within the third-generation partnership project (3GPP) framework and how our envisioned XR scenario relates to the foreseen use cases. Finally, we identify new challenges arising from integrating CC techniques into multi-user XR scenarios and propose novel solutions to address them in practice.
\end{abstract}


\maketitle

\section{Introduction}
\label{section:intro}
\IEEEPARstart{A}{n} increasingly large share of today's population is continuously connected to the virtual information world through flat-screen mobile devices, such as smartphones and tablets. The next step in the evolution of human-computer interfaces will bring forward immersive viewing experiences
%facilitated by more capable wearable gadgets.  Such immersive experiences 
that submerge users into the 3D digital world with six degrees of freedom (forward/back, up/down, left/right, yaw, pitch, roll), thus allowing them to interact with different virtual objects while remaining integrated into the real world. Although the idea of digital immersion has been around for decades, its mass adoption has been severely hampered by movement-restricting wired connections between the interfacing headsets and the external hardware. 
%This limitation, together with the low processing capabilities of the available hardware, has generally kept the immersive experience a niche concept.
However, this trend is recently starting to change as a new generation of powerful and affordable untethered headsets is introduced to the market.
% by various hardware vendors. 

Improved hardware and software capabilities are critical for supporting comfortable immersive user experiences that require spatial audio and high-definition (more than 4K per eye) video to create an accurate sensory perception of presence in a digital \rev{environment~\cite{thomas2020mpeg,taleb2022towards,chaccour2022can,boos2016flashback,chen2018virtual,bastug2017toward}}. Specifically,
% recently emerging untethered
wireless immersive applications necessitate powerful and stable
radio connectivity
%external radio connections 
%and impose stringent communication requirements such as 
with very high throughput 
%data links with 
and ultra-low latency to the processing unit,
%For example, an ultimate 3D video may require up to 3.3 Gbps of throughput, and an interactive, immersive application necessitates over-the-air latencies of around 1 ms. 
%In other words, the constraining features of human perception result in much higher importance of radio link reliability and stability for peak data rates than traditional 2D screens, 
and failure to satisfy these requirements leads to discomfort, disorientation, and nausea caused by human sensory conflict. 
Of course,
these requirements can be already satisfied in test environments (i.e., with the assumption of very large bandwidth all dedicated to XR users) using state-of-the-art technologies such as the latest third-generation partnership project (3GPP) 5G New Radio (NR) standard. However, with the growing penetration level and usage intensity of multi-user XR applications, available radio resources should be shared between the users, causing link qualities to deteriorate due to inter-user interference. Therefore, to meet future capacity demand, there is a need to further enhance the network capacity by introducing novel disruptive technology components.

%From another perspective, the software market is offering extended reality (XR) applications that engage larger numbers of people. 
%With the growing penetration levels and usage intensity, there is also a continuous need to increase network capacity with the latest available technologies. Of course, a single localized application can be easily supported today by existing wireless technologies (e.g., by the latest third-generation partnership project (3GPP) 5G New Radio (NR) standards that support data rates of multiple Gbps). However, in multi-user applications, available radio resources will be shared between the users, and the inter-stream interference will also affect the link quality. Therefore, to meet future capacity demand, there is a need to further enhance and introduce novel technologies, especially for multi-user scenarios.

%With the growing penetration levels and usage intensity, the actual deployment densities will create a situation that even the most advanced radio networks simply cannot handle. While a single localized application can be supported today by existing wireless technologies (e.g., by the latest third-generation partnership project (3GPP) 5G New Radio (NR) standards that support data rates up to multi Gbps), in multi-user applications, the interference caused by simultaneous streams will result in detrimental degradation of the radio link quality. The result is a compromised streaming or rendering quality, destroying the immersion experience for all the users in the entire network.

One feasible solution for enabling the massive use of immersive applications 
%increasing the network capacity 
is to exploit the abundant spectrum available in millimeter-wave (mmWave) bands, i.e., the Frequency Range~2 (FR2) in the 5G NR. The critical advantage of mmWave is the availability of wide bandwidth and the possibility of miniaturized antenna elements that enable multi-antenna systems with highly-directional data transmissions.
%Anticipating that, some hardware vendors have already supplemented their headsets with mmWave wireless adapters. 
However, with 
%the mmWave connectivity
this promise of high throughput connectivity also comes with specific challenges, such as
%, for example, 
volatile channel quality, random blockage effect, and complex 3D interference footprint. 
%If not handled properly, these aspects will become a critical impediment to enabling future dense and dynamic immersive networks with unconstrained user mobility. 
A significant research effort is now carried out by academia and industry to address these critical issues, using, for example, multi-point connectivity across large antenna arrays~\cite{rajatheva2020white}.

Another possible solution for enabling multi-user wireless XR applications is using novel coded caching (CC) techniques that prioritize the content relevant to wireless bottleneck areas and enable the cumulative cache memory of users in the network to be used as an additional communication \rev{resource~\cite{maddah2014fundamental}}. This solution is especially appealing as the onboard memory is becoming cheaper to implement and is available in larger quantities on modern devices.
%increasingly more devices are equipped with large-capacity memory chips. 
Caching is a well-studied concept; it has been used for a long time to place prevalent data closer to requesting users, reducing delivery time and network congestion \rev{probability~\cite{paschos2018role,sun2019communications,sun2020bandwidth,salehi2017optimality}}. However, CC extends the benefits of traditional caching \rev{techniques} by enabling a new performance gain proportional to the cumulative cache size in all users~\cite{maddah2014fundamental}. 
%Furthermore, integrating tailored multi-antenna (multicast) transmission with coded caching schemes enables the global caching gain and the spatial multiplexing gain combine in an additive manner~\cite{shariatpanahi2018physical}.
%
Interestingly, this new gain can also be combined with the multiplexing gain of \rev{multi-input single-output (MISO)~\cite{shariatpanahi2016multi,shariatpanahi2018physical,tolli2017multi,tolli2018multicast,lampiris2021resolving,mohajer2020miso} and multi-input multi-output (MIMO)~\cite{cao2017fundamental,cao2019treating,salehi2021MIMO,salehi2023multicast,NaseriTehrani2023MulticastSystems} communications} using recently introduced multi-server and multi-antenna coded caching techniques. 
%The key to achieving this combined gain is to store fragments of every file in the library in the cache memories of the users during the so-called placement phase. Then, during the delivery phase, we multicast carefully built codewords, such that every user in the target group can remove unwanted interference terms from the received signal using its cache contents.
%Indeed, similar to mmWave connectivity, achieving the performance gains of coded caching in practice, specifically in XR applications, also requires solving critical problems. %Moreover, proper frameworks are  

\rev{
Using caching techniques to improve users' quality of experience (QoE) in wireless XR applications is studied by a number of works in the literature.
From a general perspective of caching, it is shown that utilizing the storage and computing capabilities of XR mobile gadgets could effectively alleviate the traffic burden over the wireless network. Moreover, significant bandwidth and delay-reduction gains have been demonstrated~\cite{sun2019communications,yang2018communication,sun2020bandwidth,dang2019joint}.
Similarly, applying CC techniques in XR use cases has been considered by a handful of works~\cite{mahmoodi2021non,mahmoodi2022asymmetric,mahmoodi2022non,mahmoodi2023multi}. The general idea is to use two important features of wireless XR applications: 1) the content requested by users in such applications is used to reproduce their field of view (FoV) and hence is location-dependent, and 2) the size of the file library (i.e., the set of files that could be requested by users) is naturally limited. These features are employed to design new `location-dependent' CC schemes, where non-uniform portions of the cache memory are allocated to store (parts of) the content files requested in different locations within the application hall, and novel CC techniques are employed to increase the achievable data rate for the resulting cache placement.
%
%with non-uniform memory allocation to content files requested in different locations, for both single-antenna~\cite{mahmoodi2021non} and multi-antenna~\cite{mahmoodi2022asymmetric,mahmoodi2022non,mahmoodi2023multi} setups.
%
The non-uniform cache allocation allows for storing larger portions of the content files requested in locations with poor connectivity in the cache; hence, users with bad channel conditions need to download smaller data amounts from the server and do not experience excessive data delivery delays.
This results in improved QoE throughout the entire application environment by removing wireless connectivity bottlenecks.
%The goal of the non-uniform memory allocation is to remove wireless connectivity bottlenecks (as less data is requested in locations with poor connectivity) and avoid excessive communication delays in order to improve users' QoE. The result is a high-performance CC scheme that enables high-rate data delivery to multiple XR users with bounded latency throughout the entire application hall.
%where the available cache memory is allocated non-uniformly to store (parts of) the content requested in various locations in the application hall, in order to remove wireless connectivity bottlenecks and improve users' QoE. 
Due to their close alignment with the context of this paper, CC schemes of~\cite{mahmoodi2021non,mahmoodi2022asymmetric,mahmoodi2022non,mahmoodi2023multi} are reviewed in more detail in Section~\ref{sec:EnhancedXR}.\ref{section:location_dependent}. 
}

\rev{This paper takes a broader look, from both technical and standardization perspectives, at the benefits and problems of applying CC techniques, especially in the context of multi-antenna communications, to address
%This paper specifically explores how multi-antenna coded caching techniques could help address 
critical communications bottlenecks of multi-user XR applications.}
%In general, coded caching techniques can be used whenever a set of users in the network request data from a limited library. 
In this regard, we first introduce our envisioned cache-aided multi-user XR setup and study recent multi-antenna CC advancements. Then, we review how standardization bodies, especially 3GPP, consider XR requirements in their specification studies and explore how our envisioned XR scenario relates to their foreseen use cases. 
Finally, we discuss in more detail the challenges arising while integrating CC techniques into multi-user XR applications and propose novel solutions to address the identified issues. %The ultimate goal is to pave the way for coded caching to be considered as an improved connectivity enabler for XR applications in future standardization. 

%\todo[inline]{We need to improve the novelty part. Ideas: 1) How to apply CC for XR. What are the impacts and requirements for doing that. Impacts: receiver impact, data delivery/application awareness, rendering model, data model (zoning),}

%Finally, we discuss in more detail how coded caching techniques can be tailored for XR applications and propose open questions to be addressed by the research community.

%\vspace{-5pt}
\subsection{The envisioned XR scenario}
\label{section:xr_scenario}
\begin{figure}[t]
\begin{center}
    \includegraphics[width=\columnwidth]{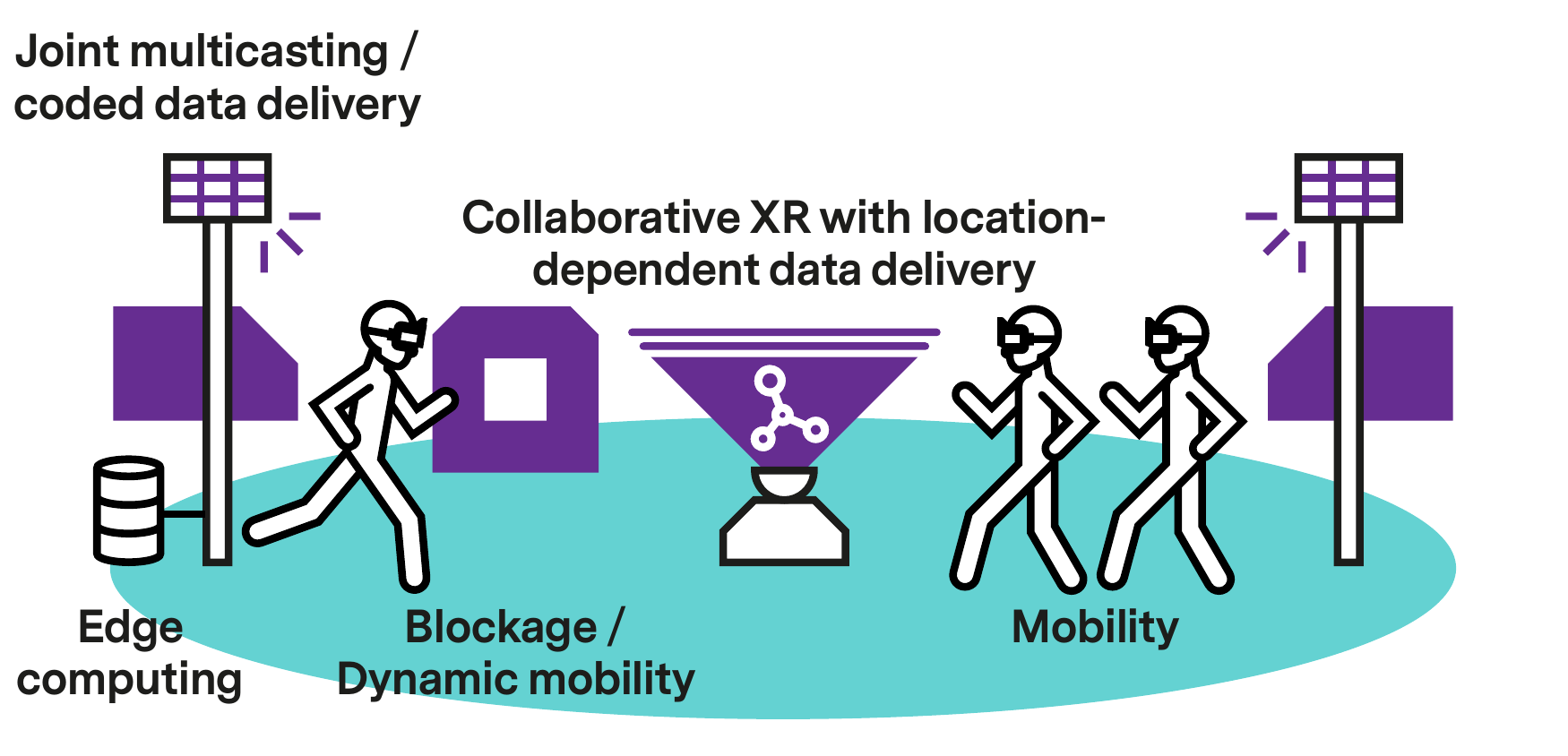}
    \caption{Immersive viewing scenario with coded caching.}
    %\vspace{-15pt}
    \label{fig:immersive_vieweing}
\end{center}
\end{figure}

We envision an XR or hyper-reality environment where a large group of users is submerged into a network-based immersive application. The use case of the setup can be, for example, educational, industrial, gaming, defense, or social networking~\cite{3GPPRef}. The XR application runs on high-end eyewear gadgets that require heavy multimedia traffic and are bound to guarantee a required level of user QoE within the operating theatre. A general illustration of such a scenario is shown in Figure~\ref{fig:immersive_vieweing}.

Increasing computation capabilities of modern chipsets allow more and more processing to be handled locally within high-end XR devices. However, with the stringent requirements of immersive applications and the form factor constraints of XR devices limiting the allowed heat dissipation level, full local processing of XR applications is still unattainable.
%make the prospect for fully local processing uncertain. 
This has led to the development of load-splitting models that enable major parts of the computation to \rev{take place} in edge servers while offloading delay-sensitive final refinements to be handled locally at XR devices~\cite{3GPPRef}. Such load-splitting models rely heavily on a reliable and fast communication channel between the edge and XR devices, as the servers should constantly be informed of the application environment and transfer their generated processing and rendering outputs to the end users~\cite{3GPPRef}.
In this regard, in our model, we assume that the \rev{transceivers on interfacing eyewear devices support high throughput demands, e.g. by operating on mmWave bands,}
%
%\revB{While the growing computation capability of modern chipsets used in XR devices allows a larger portion of the processing to be done locally, the challenging amount of calculations as well as the form-factor constraints of wearable devices still makes it impossible to do the entire computation locally. This leads us to innovative load-splitting solutions with the edge servers, where part of the process is done in the edge, the result is delivered to the end-users through high-throughput links, and final corrections are made locally.}
%As computation capabilities and form-factor constraints of wearable devices may not permit entirely local content processing, network edge infrastructure is used to assist with the rendering. The resulting content is then delivered to the users via high-throughput wireless links.
%, and at the same time, fully network-based solutions may not be possible any time soon (network delays are a few orders higher than acceptable values for immersive experience [Han2017]), the only viable option to provide continuous quality of immersive experience remains in the use of the network edge infrastructures and/or in disseminating some of the data across the network via multipoint multicast beamforming of network coded content as well as employing proximate device-to-device connections (D2D) for direct content exchange.
%The 
%interfacing eyewear devices \rev{have transceivers that support high throughput demands, e.g. by operating on mmWave bands,} 
and remain connected to the network, possibly through multiple transmit-receive points (TRPs) that enable reliable communication via multi-connectivity schemes to provide improved resilience to random blockage and increased coverage and capacity in the application area~\cite{Kumar2022Latency-AwareConnectivity}.

%\rev{
%
%\textbf{maybe we should move this paragraph from here... or at least make the explanation briefer and refer to the next sections. \\
%A good part of the content is cacheable... as an example, consider a gaming scenario where the content can be split into static and dynamic parts as shown in the figure...
%}

%\textbf{Maybe more discussion on why CC fits XR.}

%}

\begin{figure}[t]
     \centering
      \includegraphics[width=0.85\columnwidth]{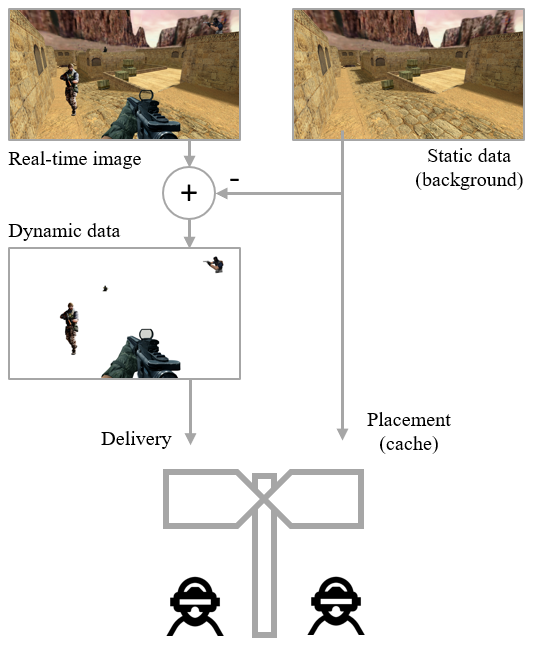}
        %\vspace{-3mm}
        \caption{Data decomposition into static and dynamic parts 
        (screenshots are from the counter-strike game).
        }
        \label{fig:static_dynamic_decomposition}
        %\vspace{-2mm}
\end{figure}

We also assume that multimedia consumers (i.e., network users) are scattered across the application area and can move freely, and their streamed data is unique and highly location- and time-dependent~\cite{mahmoodi2021non}.
%\rev{
%Depending on the underlying model for 3D graphic rendering and content splitting, coded caching techniques could be used in various ways to enhance data delivery in the considered XR setup. While this is elaborated in Section ..., as a simple example, consider a collaborative XR scenario where multiple users are engaged in an interactive gaming application within a closed hall. Indeed, such a gaming scenario requires instant delivery of frequently changing data elements. This so-called \emph{dynamic} data part includes, for example, the data related to the movements and actions of other users. 
%}
Like any other interactive scenario, our envisioned application also requires instant delivery of frequently changing data elements. This so-called \emph{dynamic} data part includes, for example, the data related to the movements and actions of other users in a gaming application. 
However, a notably large part of the delivered content, e.g., the data required for rendering the background scenery of the user's FoV, is \emph{static} (i.e., non-interactive) and can be cached beforehand when favorable radio channel conditions and excess communication resources  are available (see Figure~\ref{fig:static_dynamic_decomposition}).\footnote{
In typical virtual gaming applications, even the dynamic part of the FoV can be described by smaller, cacheable elements. For example, a moving person in an XR game can be represented as a superposition of multiple well-defined geometrical shapes covered with various textures and possibly overlaid by the avatar of the corresponding player. All these elements (geometrical shapes, textures, and avatars) are cacheable and can be efficiently delivered using coded caching techniques (i.e., by storing part of the elements and multicasting the rest). Of course, one would still need to transmit control/instruction data describing how to reconstruct the dynamic part from both the cached elements and the multicast data. More details can be found in~\cite{thomas2020mpeg,boos2016flashback}.
%The separation of static and dynamic content for XR applications is not unique to this paper. Check~\cite{thomas2020mpeg} for another reference.
}
%An example of data composition to static and dynamic parts is provided in Fig.~\ref{}.
%Like any other interactive scenario, our envisioned application also requires instant delivery of frequently changing data elements. This dynamic data part includes, for example, the data related to the movements and actions of other users in a gaming application. However, a notably large part of the delivered content (e.g., the data required for rendering the background scenery of the user's field of view) is static and can be cached beforehand when favorable radio channel conditions and excess communication capacity exist. 
This feasibility of caching, together with the limited number of cachable files, provides the opportunity for efficient use of pooled memory resources through intelligent coded caching mechanisms~\cite{mahmoodi2021non,mahmoodi2022non,mahmoodi2022asymmetric}. The result is the possibility of delivering high-throughput, low-latency data traffic while providing high stability and reliability of radio connections for a truly immersive experience. 
%Modern mobile devices continue to grow the amounts of available storage, making coded caching especially beneficial in use cases where the popularity of limited and location-dependent content becomes much higher than in any traditional network. 
%When user devices are in the proximity of a certain TRP, which acts as a ``data shower'' or a ``data kiosk'', line-of-sight (LoS) MIMO links enable dynamic (re-)placement of cached content at the very low cost of radio \rev{and energy} resources. The phase of dynamic content (re-)placement repeats itself with more relevant multimedia content (e.g., the content that is more likely to be requested following \rev{one's} movement direction) when users approach other TRPs.

%\todo[inline]{the following paragraph is a copy of the AoF-NSF application. It would need some polishing to better fit the current text flow.}

\section{Coded Caching (CC)}
\label{section:codedcachingreview}
The pioneering work in~\cite{maddah2014fundamental} introduced a novel \emph{coded caching} scheme where instead of merely replicating high-popularity content near (or at) end-users, fragments of the contents were spread across different cache memories throughout the network. CC works in two phases, \emph{placement} and \emph{delivery}. During the placement phase, content files are split into smaller parts, and these parts are stored in the cache memories of different users. This phase is performed, for example, when the network traffic is low (to avoid congestion) or when the users are close to TRPs (to minimize the transmission time and energy expenditure). Then, during the delivery phase, after the network users reveal their requests, several codewords are built and each codeword is multicast to a subset of target users. 

%Placement and delivery phases are designed to ensure that after multicasting each codeword, every target user can remove undesired terms from the received signal using its cache contents and get (parts of) its requested data interference-free.
%\todo[inline]{we can put a dedicated paragraph about the energy expenditure of the placement phase here: data is delivered in an energy-efficient manner, as when the users are close to TRPs, the datarate is high and the transmission time is small. Also, even though part of the data might never be used, some other parts may be used multiple times.}

\begin{figure}[t]
\begin{center}
    \includegraphics[width=0.9\columnwidth]{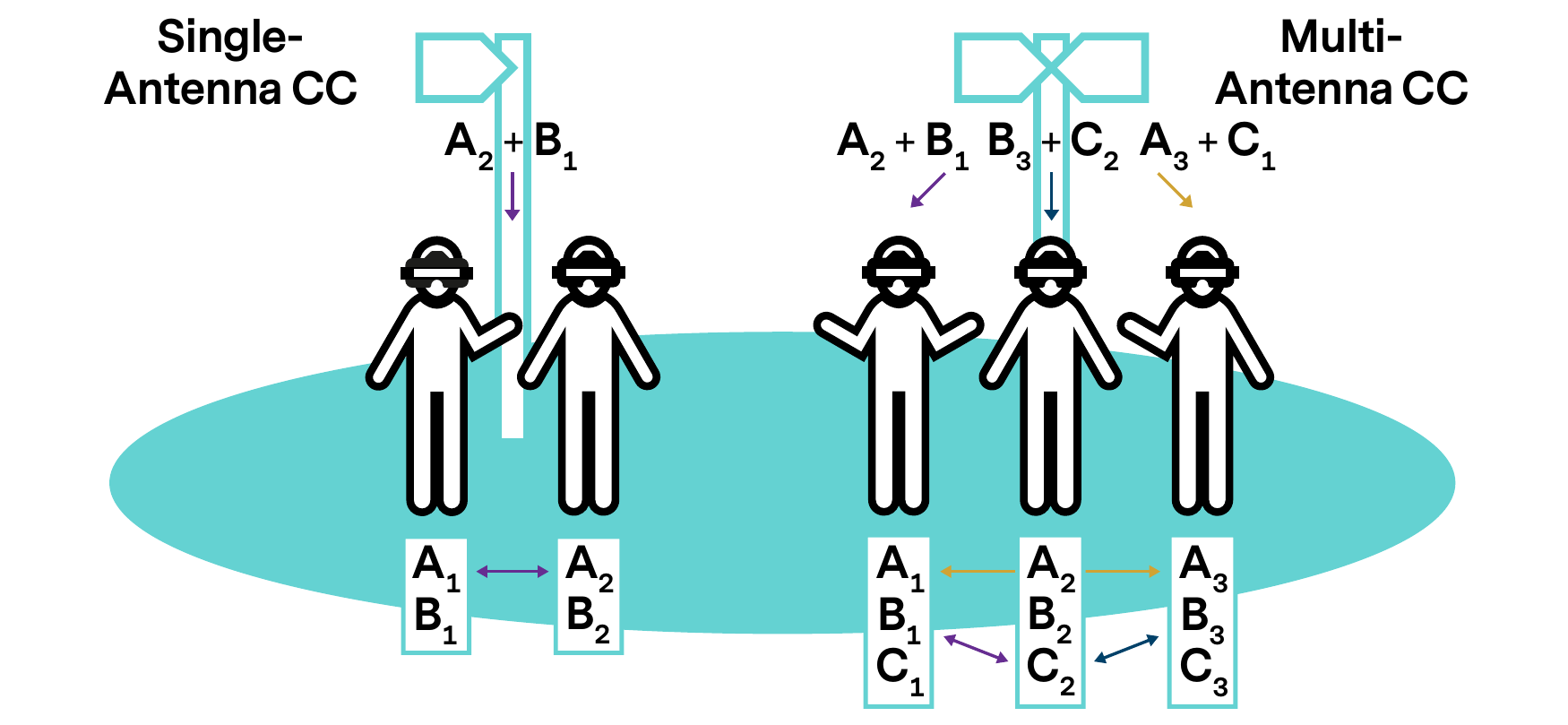}
    %\vspace{-30pt}
    \caption{Single- and multi-antenna coded caching}
    \label{fig:cc_illustration}
    \vspace{-10pt}
\end{center}
\end{figure}

A simple illustration of a CC-aided operation with a single-antenna transmitter is shown on the left-hand side of Figure~\ref{fig:cc_illustration}. In this simplified scenario, two users, each with cache memory large enough to store one of the files $A$ and $B$, request files from a single-antenna server over a shared link. 
%Both users will request one of the files $A$ and $B$, 
There is no prior information about the request probabilities. 
It can be shown that with a classic caching solution (i.e., no coded placement or delivery), the worst-case load on the shared link cannot be smaller than the size of one file.
%Using a classic caching solution, we have to cache files randomly; for example, to store $A$ at user one (left) and $B$ at user two (right). It can be shown that with such a cache placement, depending on the users' actual requests, 
%we need to send one file over the shared link by average. 
However, using CC, we can halve this worst-case load %by a factor of two. 
%This is done 
by allowing each transmission to serve two users simultaneously. For the example network, this is achieved by splitting each file into two equal-sized parts, and caching $A_1$ and $B_1$ at user one and $A_2$ and $B_2$ at user two. Then, for example, if users one and two request files $A$ and $B$, respectively, we simply transmit $A_2 \oplus B_1$, where $\oplus$ denotes the XOR operation over the finite field. This way, users one and two can remove undesired terms $B_1$ and $A_2$ from the received signal using their cached contents and decode $A_2$ and $B_1$, respectively. %interference-free. As it has $A_1$ already cached in the memory, it then has its requested file $A$ completely. Similarly, user two can remove $A_2$ from the received signal to have its requested file $B$. 
%The 
%file parts $A_2$ and $B_1$ each have the size of half of one file, and the 
%XOR operation does not change the size, and hence, the transmitted data size is halved compared with classic caching. 
%By checking all possible request patterns, it can be shown that CC enables a performance boost by a factor of two for this example network.
%\rev{Going through all possible request patterns, we observe that CC reduces the worst-case load on the shared link by a factor of two.}
For better clarification, in Table~\ref{tab:cc_classic_comp}, we have compared both classic and coded caching schemes for this example network, considering all possible request patterns.
In general, using CC, the number of users served with each transmission (hence, the reduction in the worst-case load) scales with
%coded caching enables boosting the performance \rev{(in terms of reducing the worst-case load on the shared link)} by a multiplicative factor proportional to 
the cumulative cache size in the network~\cite{maddah2014fundamental}.
\rev{In mathematical terms, if each user can cache a portion $\gamma$ of the entire library and there exist $K$ users in the network, we can serve $t+1$ users in each transmission, where $t = K\gamma$ is called the \emph{coded caching gain}}.

\begin{table*}[ht]
    \centering
    \small{
    \begin{tabular}{|c|c||c|c|c|c||c|c|c|c||}
         %\cline{3-10}
         \multicolumn{2}{c||}{} & \multicolumn{4}{c||}{\textbf{Classic Caching}} & \multicolumn{4}{c|}{\textbf{Coded Caching}} \\
         \hline
         \multicolumn{2}{|c||}{requests} & \multicolumn{2}{c|}{cache contents} & \multirow{2}{*}{transmitted data} & \multirow{2}{*}{link load} & \multicolumn{2}{c|}{cache contents} & \multirow{2}{*}{transmitted data} & \multirow{2}{*}{link load} \\
         \cline{1-4}\cline{7-8}
         user 1 & user 2 & user 1 & user 2 & & & user 1 & user 2 & & \\
         \hline
         $A$ & $B$ & \multirow{4}{*}{$A$} & \multirow{4}{*}{$B$} & - & 0 & \multirow{4}{*}{$A_1$, $B_1$} & \multirow{4}{*}{$A_2$, $B_2$} & $A_2 \oplus B_1$ & $0.5$ \\
         \cline{1-2}\cline{5-6}\cline{9-10}
         $B$ & $A$ & & & $B$, $A$ & 2 & & & $B_2 \oplus A_1$ & $0.5$ \\
         \cline{1-2}\cline{5-6}\cline{9-10}
         $A$ & $A$ & & & $A$ & 1 & & & $A_2 \oplus A_1$ & $0.5$ \\
         \cline{1-2}\cline{5-6}\cline{9-10}
         $B$ & $B$ & & & $B$ & 1 & & & $B_2 \oplus B_1$ & $0.5$ \\
         \hline
         \multicolumn{2}{c||}{} & \multicolumn{3}{c|}{\textbf{Average Link Load:}} & \textbf{1} & \multicolumn{3}{c|}{\textbf{Average Link Load:}} & $\mathbf{0.5}$ \\
         %\cline{3-10}
    \end{tabular}
    }
   \caption{Request patterns, transmitted data, and link loads in classic and coded caching, for the example network}
    \label{tab:cc_classic_comp}
    %\vspace{-1mm}
\end{table*}

%\todo[inline]{Reduce the above paragraph: maybe putting a table helps.}

%In other words, each content file is first split into a number of equal-sized smaller parts, which are stored in different cache memories throughout the network following a \emph{cache placement} algorithm. Then, during the so-called delivery phase, based on the content requests received from network users, a number of carefully created codewords are broadcast to various groups of users. The codewords are created such that within each target user group, every user is able to remove unwanted terms from the received signal and decode parts of its requested content file interference-free. 

%In general, coded caching enables every transmitted data bit to contain useful data for all the target users, and hence, the effective rate of the channel is multiplied by the number of target user. As shown in~\cite{maddah2014fundamental}, with CC, the maximum number of target users is proportional to the cumulative cache size in the network, and hence, CC improves the effective rate by a multiplicative factor scaling with both the number of users and their individual cache sizes. This scaling factor was later shown to be optimal under simple assumptions of uncoded data placement and single-shot delivery (i.e., when the received signal after each transmission can be decoded without relying on other transmissions).

%\vspace{-5pt}
%\subsection{Multi-antenna coded caching}

\begin{figure*}[t]
    \begin{subfigure}{\textwidth}
        \centering
        \includegraphics[width = 0.9\textwidth]{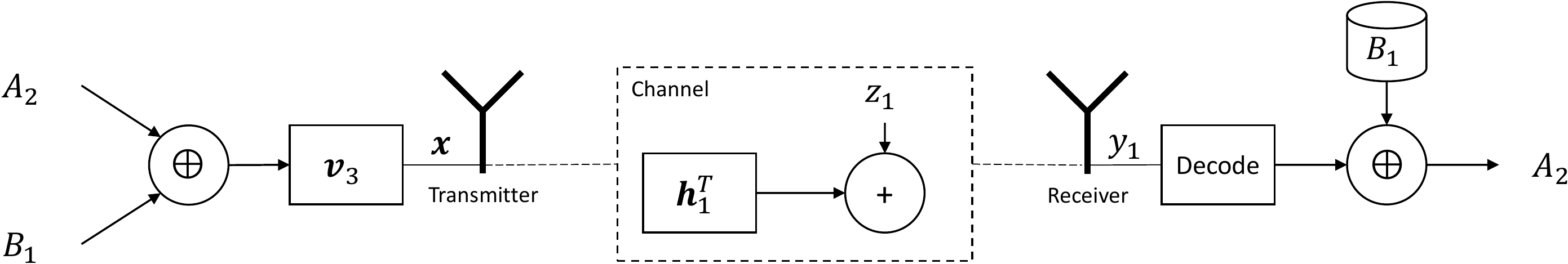}
        %\vspace{5pt}
        %\caption{Bit-level interference cancellation}
        %\label{fig:subfig_bitlevel}
    \end{subfigure}
    \vspace{25pt}
    \begin{subfigure}{\textwidth}
        \centering
        \includegraphics[width = 0.9\textwidth]{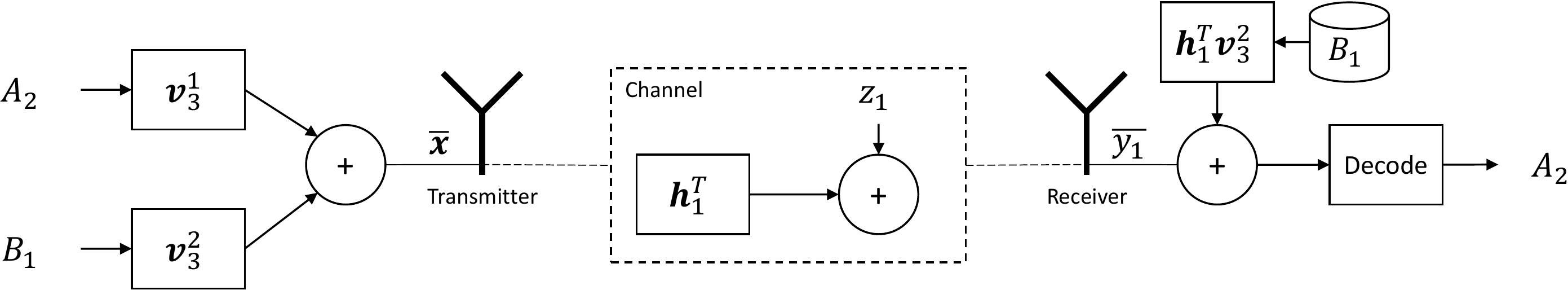}
        %\caption{Signal-level interference cancellation}
        %\label{fig:subfig_siglevel}
    \end{subfigure}
    \vspace{-10pt}
    \caption{Bit-level (top) and signal-level (below) schemes %Figure from~\cite{salehi2021MIMO}
    %\vspace{-5pt}
    }
    \label{fig:bit_sig_level}
\end{figure*}

Following the original CC scheme in~\cite{maddah2014fundamental}, many ensuing works in the literature applied its core idea to more diverse network setups. An important direction
%, initiated by~\cite{shariatpanahi2018physical}, 
was to apply CC techniques in MISO setups,
%where 
%
%a multi-antenna server communicates with multiple single-antenna users~\cite{shariatpanahi2018physical}. It is well-known that, using appropriate transmit precoders (i.e., beamformers) that \rev{suppress} undesired terms at each receiver, MISO setups provide a \emph{multiplexing gain}; i.e., with $L$ adequately-spaced antennas, the server can transmit parallel data streams to $L$ users simultaneously.
%
%In a general multi-server scenario where multiple transmitting servers cooperatively send data to a set of requesting users over linear, independent communication channels, it is well-known that one can simultaneously send data to a subset of users interference-free.
%This \emph{multiplexing gain} is enabled by appropriate transmit precoders (i.e., beamformers) that null out (or suppress) undesired terms at each receiver. 
%Of course, the value of the multiplexing gain (i.e., the size of the subset of users being served simultaneously) cannot be larger than the number of servers.
%The work in~\cite{shariatpanahi2018physical} %extended the multi-server scenario by considering cache-enabled users and 
%revealed the exciting result that in a MISO setup with cache-enabled users, 
revealing the exciting result that the CC and spatial multiplexing gains are additive.
\rev{In fact, with coded caching gain $t$, if the spatial multiplexing gain of $L$ is attainable by the transmitter,\footnote{\rev{
The spatial multiplexing gain refers to the total number of parallel streams (across multiple users) that can be handled by a single access point, and its value is upper-bounded by the number of antennas at that access point. While with the recent emergence of communications in higher frequencies (e.g., in mmWave bands), it has become possible to employ larger antenna arrays ($>100$) at both transmitters and receivers, the attainable multiplexing gain is more constrained due to practical limitations in the number of RF chains and available baseband processing and transmit power, as well as severely constrained pilot resources for channel sounding reference symbols~\cite{3GPP_36.101,3GPP_protocol,5gprotocol_book}. Of course, as shown in~\cite{tolli2017multi}, MISO CC techniques can significantly benefit from improved multicast beamforming gains when the spatial multiplexing gain is smaller than the antenna count.
}} the total number of users served in each transmission can reach $t+L$. In~\cite{lampiris2021resolving}, it is shown that this number is optimal under the conditions of uncoded cache placement and single-shot data delivery (i.e. when the decoding process of one transmission does not depend on other transmissions). As a clarifying}
%In other words, it was shown that with appropriate coded caching techniques, the number of target users could be larger than the antenna count, as part of the interference at each user could be removed using its cache contents.
%
%Following a similar approach to~\cite{shariatpanahi2016multi}, applying coded caching techniques in multi-input single-output (MISO) setups is later considered in~\cite{shariatpanahi2018physical}. The extension is straightforward; each antenna element at the transmitter is analogous to an individual server in the multi-antenna setup, and transmit precoders for nulling out (part of) the interference are replaced by zero-forcing beamformers.
%
%A small example of CC-aided operation in a 
example, consider the
MISO setup in the right-hand side of Figure~\ref{fig:cc_illustration}, where three single-antenna users, each with a cache memory large enough to store one file, request data from a \rev{ multi-antenna server that can attain a spatial multiplexing gain of $L=2$}. Requests are made from a library of three files $A$, $B$, and $C$ \rev{(i.e., $\gamma = 1/3$ and $t=1$)}, and there is no prior knowledge of the requests. For this network, by splitting each file into three equal-sized parts and caching them as shown in Figure~\ref{fig:cc_illustration}, we can serve all \rev{$t+L=3$} users simultaneously (while without CC, at most \rev{$L=2$} users can be served). To achieve this additional gain, we create an appropriate XOR codeword for every group of users of size two, and design multicast beamformers to suppress the interference caused by each of these group-specific codewords at the other user not belonging to that group.
For example, assuming users~1-3 have requested files $A$, $B$, and $C$, respectively, the transmission vector for this network is built as
\begin{equation}
    \label{eq:trans_vector_bit_level}
    \Bx = (A_2 \oplus B_1) \Bv_3 + (A_3 \oplus C_1) \Bv_2 + (B_3 \oplus C_2) \Bv_1 ,
\end{equation}
where $\Bv_k$ is the beamforming vector designed to suppress interference at user~$k$.\footnote{As in conventional MISO systems without coded caching, various strategies can be used to design beamformers, and they affect the system performance depending on the operating signal-to-noise ratio (SNR). A detailed discussion on beamforming for MISO coded caching schemes is provided in~\cite{tolli2017multi}.}
Let us review the decoding process at user~1, after the transmission of $\Bx$. Denoting the channel vector of this user and the additive noise at its receiver with $\Bh_1$ and $z_1$, respectively, we can model the received signal as $y_1 = \Bh_1^T \Bx + z_1$, i.e.,
\begin{equation}
\label{eq:rec_signal_bit_level}
    \begin{aligned}
        y_1 = &(A_2 \oplus B_1) \Bh_1^T\Bv_3 + (A_3 \oplus C_1) \Bh_1^T\Bv_2 \\
        & + \underline{(B_3 \oplus C_2) \Bh_1^T\Bv_1} + z_1 .
    \end{aligned}
\end{equation}
According to the definition, the interference caused by the underlined term in~\eqref{eq:rec_signal_bit_level} is suppressed by beamforming vector $\Bv_1$. Hence, user~1 can decode the two codewords $(A_2 \oplus B_1)$ and $(A_3 \oplus C_1)$ from $y_1$, using, e.g., a successive interference cancellation (SIC) receiver. Finally, as user~1 has $B_1$ and $C_1$ in its cache (see Figure~\ref{fig:cc_illustration}), it can extract its desired terms $A_2$ and $A_3$ out of the codewords readily using an XOR operation.

\subsection{Coded caching with signal-level interference cancellation}
%One major bottleneck in the practical implementation of coded caching techniques is the subpacketization issue. 
In general, CC techniques rely on spreading content fragments over the cache memories throughout the network. These small fragments are usually called \emph{subpackets} in the literature, and the process of splitting content files into subpackets is known as \emph{subpacketization}~\cite{lampiris2018adding}. 
%The number of fragments required for each content file is called \emph{subpacketization}. 
For example, for the single- and multi-antenna setups in Figure~\ref{fig:cc_illustration}, content files are split into two and three subpackets, respectively. The problem is that, for both baseline single-antenna~\cite{maddah2014fundamental} and multi-antenna~\cite{shariatpanahi2018physical} CC schemes, the number of subpackets (i.e., the subpacketization value) grows exponentially with the number of users, severely limiting 
%The problem is even worse for the baseline multi-antenna scheme in~\cite{shariatpanahi2018physical}, as its subpacketization requirement is larger than that of~\cite{maddah2014fundamental} by a multiplicative factor which itself grows exponentially with the user count. 
%As thoroughly studied in~\cite{lampiris2018adding}, such an exponential growth severely limits 
the real achievable CC gain in practice~\cite{lampiris2018adding}.
%Consequently,
%reducing subpacketization with no or minor impact on the achievable gain has been thoroughly investigated in the literature.
%for both single- and multi-antenna setups. 

While single-antenna setups are not flexible for reducing the subpacketization value~\cite{yan2018placement}, multi-antenna setups can work with modest subpacketization values, even smaller than their comparable single-antenna counterparts~\cite{lampiris2018adding}. This is enabled by a new cache-aided interference cancellation mechanism, where unwanted terms are regenerated from the local memory and removed \textit{before} the received signal is decoded at the receiver.
We use the term \emph{signal-level} approach for this new mechanism and denote the classic cache-aided interference cancellation mechanism (after decoding at the receiver) as the \emph{bit-level} approach.

For clarification, let us review how the signal-level approach could be applied to the MISO setup in Figure~\ref{fig:cc_illustration}. For this network, instead of transmitting $\Bx$ in~\eqref{eq:trans_vector_bit_level}, one can use
\begin{equation}
    \label{eq:trans_vector_signal_level}
    \Bar{\Bx} = A_2 \Bv_3^1 + B_1 \Bv_3^2 + A_3 \Bv_2^1 + C_1 \Bv_2^2 + B_3 \Bv_1^1 + C_2 \Bv_1^2
\end{equation}
to deliver the same data terms to every user. In~\eqref{eq:trans_vector_signal_level}, $\Bv_k^1$ and $\Bv_k^2$ denote two (possibly different) beamformers, both suppressing data at user~$k$.
For clarification, let us review the decoding process at user~1. Following the same discussions as before, this user receives
\begin{equation}
    \begin{aligned}
        \Bar{y}_1 = &A_2 \Bh_1^T\Bv_3^1 + B_1 \Bh_1^T\Bv_3^2 + A_3 \Bh_1^T\Bv_2^1 \\
        &+ C_1 \Bh_1^T\Bv_2^2 + \underline{B_3 \Bh_1^T\Bv_1^1} + \underline{C_2 \Bh_1^T\Bv_1^2} + z_1 ,
    \end{aligned}
\end{equation}
where the interference from underlined terms is suppressed by beamformer vectors $\Bv_1^1$ and $\Bv_1^2$. Now, user~1 has to first regenerate the remaining interference terms $B_1 \Bh_1^T\Bv_3^2$ and $C_1 \Bh_1^T\Bv_2^2$ using its cache contents\footnote{The effective channel coefficients $\Bh_1^T\Bv_3^2$ and $\Bh_1^T\Bv_2^2$ can be estimated, for example, from demodulation reference signal (DMRS) pilots.} and remove them in the signal domain \rev{from the received signal before it can decode its desired terms $A_2$ and $A_3$.}
%, similarly to  SIC receiver structure. 
This contrasts the bit-level approach where the cache contents were used after the received signal was decoded at the receiver.
A graphical comparison of bit-level and signal-level approaches for the considered example network is provided in Figure~\ref{fig:bit_sig_level}. Note that in this figure, only the encoding process of $A_2$ and $B_1$ and the decoding process of $A_2$ at user~1 are shown.

%To see how the decoding and encoding processes differ for $\Bx$ (bit-level) and $\Bar{\Bx}$ (signal-level), in Figure~\ref{fig:bit_sig_level}, we have depicted the encoding process of $A_2$ and $B_1$ and the decoding process of $A_2$ at user~1, as each transmission vector is used for data delivery.

%
%In Figure~\ref{fig:bit_sig_level}, we have compared the two approaches for the MISO setup in Figure~\ref{fig:cc_illustration}. The figure shows only the codeword generated for users one and two and the decoding process at user one, assuming these two users \rev{have requested} files $A$ and $B$, respectively. 

%As can be seen, \rev{with} the bit-level approach, the two terms $A_2$ and $B_1$ are first encoded with an XOR operation in the finite field, and next, a single beamformer $\mathbf{v}_3$ is used to \rev{suppress} the interference at user three. Then, at the receiver of user one, we first decode the received signal before using the cached term $B_1$ to recover the requested term $A_2$. However, \rev{with} the signal-level approach, we first use two beamformers $\mathbf{v}_3^1$ and $\mathbf{v}_3^2$, and then, add the resulting terms in the signal domain. Now, for decoding $A_2$ at user one, we should first \revB{regenerate and} remove the interference \rev{term $B_1 \Bh_1^T \Bv_3^2$, where $\Bh_k$ is the channel vector of user~$k$,} in the signal domain.
%\rev{and} before decoding the received signal. 

Although the signal-level approach incurs a noticeable performance loss at the finite-SNR regime due to its inferior multicasting gain compared with the bit-level approach~\cite{salehi2019subpacketization,salehi2022multi}, \rev{it provides great flexibility in addressing many practical bottlenecks of CC schemes, and as a result, is thoroughly studied in the literature. For the sake of brevity, we briefly review a few notable research directions here:

\noindent\ding{227} With signal-level interference cancellation, it is possible to reduce the subpacketization requirement while serving the same number of $t+L$ users as bit-level schemes in each transmission~\cite{lampiris2018adding}. It is even shown that the exponential subpacketization growth (with respect to the network size) in bit-level schemes can be replaced by linear scaling in MISO networks with a large spatial multiplexing gain at the transmitter~\cite{salehi2020lowcomplexity}.

\noindent\ding{227} With signal-level interference cancellation, the complex multi-group multicast beamformer design of bit-level schemes~\cite{tolli2017multi} could be replaced with a much simpler multi-user unicast beamformer design~\cite{salehi2020lowcomplexity,salehi2021low}. This allows us to design CC schemes that are applicable to very large networks and perform well in the finite-SNR regime.

\noindent\ding{227} Signal-level interference cancellation allows designing CC schemes for dynamic networks where the users can join and leave the network freely~\cite{salehi2021low,abolpour2022coded,abolpour2023cache}. Such dynamic schemes are based on the shared-cache model~\cite{parrinello2019fundamental,parrinello2020extending}, and are described in more detail in Section~\ref{sec:EnhancedXR}.\ref{section:dynamic_cc}.

\noindent\ding{227} In~\cite{salehi2021MIMO}, a signal-level CC scheme is introduced for MIMO setups (with multiple antennas at both the transmitter and receivers) that provides a multiplicative boost in the CC gain compared to MISO setups with a very low subpacketization overhead.

We should note that in addition to performance and flexibility, bit- and signal-level CC schemes are also different in signaling requirements and the way they affect the physical layer. This is elaborated on in Section~\ref{sec:EnhancedXR}.\ref{section:signal_bit_level_selection}.
}

\section{Extended Reality and Standardization}
\label{section:3GPP}
XR is expected to be one of the key 6G drivers~\cite{rajatheva2020white}. Various industrial, educational, gaming, and social networking wireless XR applications will emerge in the coming years, and they will push the capabilities of the underlying communication infrastructure. Accordingly, XR-related technical discussions are already in progress in various standardization bodies such as 3GPP.
%, the results of such discussions are made available publicly through various technical reports (TR) such as~\cite{3GPPRef}.
%
%With the expected emergence of wireless XR applications, its related technical discussions are also in progress in various standardization bodies such as 3GPP. 
To date, classifications and requirements of XR applications and their interconnection with 3GPP terms and standards are studied, and the results are made available in various technical reports (TR).
% such as~\cite{3GPPRef}. Moreover, 
%
More precisely, XR and cloud gaming (CG) evaluation methodology and performance were studied in 3GPP Release~16, with TR~38.838~\cite{3GPPRef38838} providing comprehensive performance analysis for XR and CG in relevant 5G NR deployment scenarios. This work continued in Release~17 within new study items with a focus on improvements for XR. 
The results of these study items are reported, e.g., in TR~38.835~\cite{3GPPRef38835}. 
Finally, in Release~18, NR enhancements for facilitating XR are standardized. This includes, e.g., radio access networks XR awareness operations, and XR-specific enhancements for power saving and capacity~\cite{3GPPRefRP223502}.

%which includes, e.g., application awareness operations and aspects related to XR-specific power consumption and capacity~\cite{3GPPRefRP213658}. 

In this section, we briefly overview the comprehensive report in TR 26.928~\cite{3GPPRef}, which provides a general introduction to XR terms and definitions, classifies XR device form factors and use cases, and clarifies the technical requirements of various XR applications.
%and whether they can be addressed in the current framework or need new items to be added in future releases. The discussions in this section clarify how our envisioned XR scenario lies within 3GPP use cases. 
Then, in Section~\ref{sec:EnhancedXR}, we specifically address practical implementation and specification challenges for enabling CC techniques in multi-user XR environments. 
%for their implementation in XR should be solved before their practical implementation. We also propose new tricks and ideas to address these problems.

%\subsection{XR device form factors}
%XR device classification is currently done in~\cite{3GPPRef} according to the device physical form (flat screen, mounted headset, and eyewear), process location (external or inbuilt), and the cellular communication module location (external or inbuilt). These properties affect the amount of the power that can be safely dissipated, and hence, the achievable communication and processing power in the device. An schematic description of the XR device classification is shown in Figure~\ref{fig:form_factors}.

%\revA{
%\vspace{-5pt}
\subsection{XR form factors and use cases}
\begin{figure}[t]
\begin{center}
    \includegraphics[width=0.9\columnwidth]{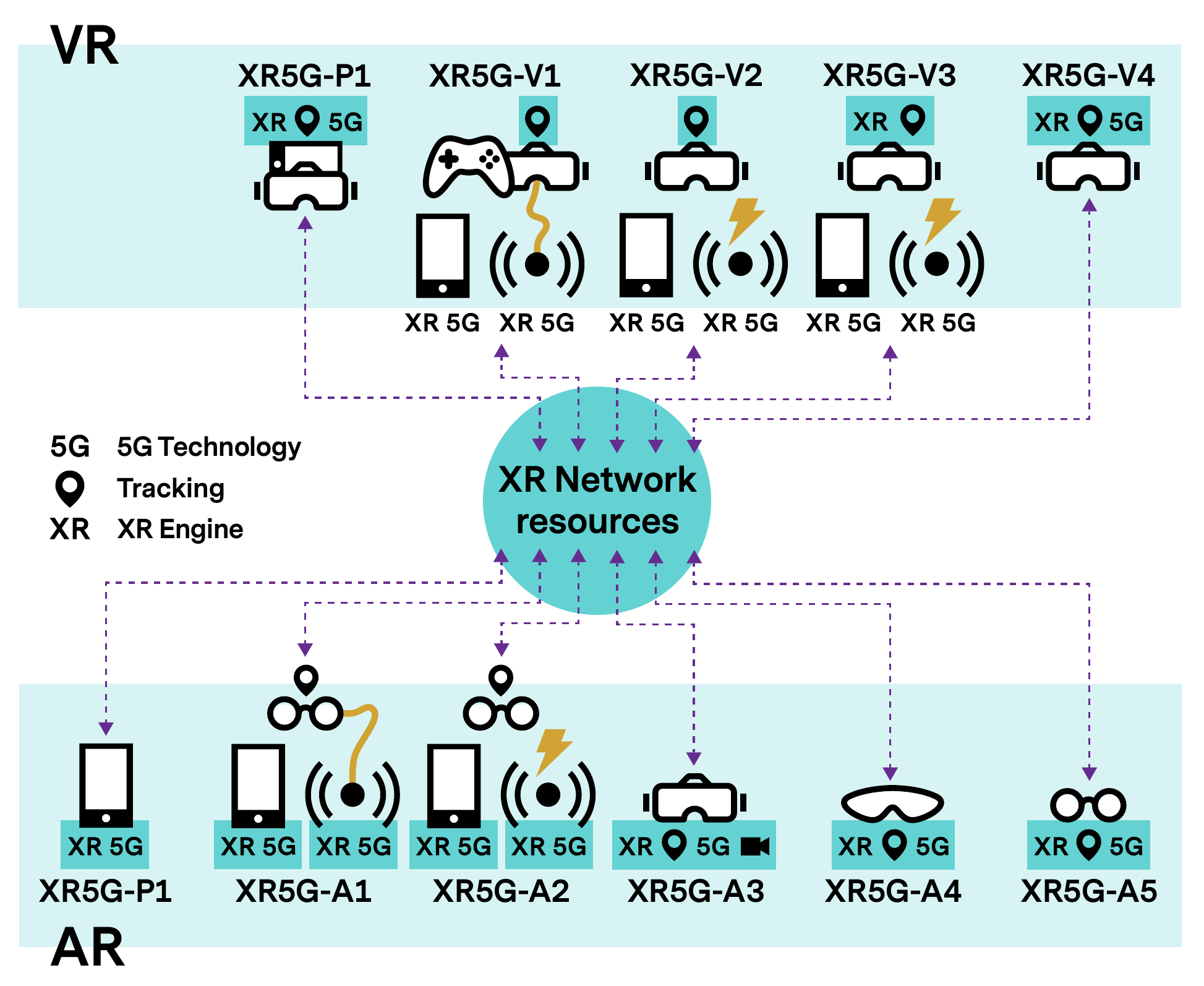}
    \caption{Device form factors as categorized in~\cite{3GPPRef}}
    \label{fig:form_factors}
    %\vspace{-15pt}
\end{center}
\end{figure}

%and use cases}
%\noindent\textbf{Device form factors.} 
XR device classification is done according to the physical form (flat screen, mounted headset, eyewear), processing unit location (external or inbuilt), and cellular communication unit location (external or inbuilt). These properties affect the amount of power that can be safely dissipated and hence, the achievable communication and processing power in the device. A schematic description of the XR device classification, as categorized in~\cite{3GPPRef}, is shown in Figure~\ref{fig:form_factors}.
\rev{
In this classification, depending on the application type, which can be either augmented reality (AR) or virtual reality (VR), XR devices are put into two broad categories. In both categories, it is possible to use the smartphone screen as the main display (denoted by XR5G-P1 in Figure~\ref{fig:form_factors}). Otherwise, if a head mount device (HMD) is used as the display to enhance the user experience, the categorization is based on: 1) where the required XR processing at the user side is performed and 2) where the 5G transceiver is located. Both processing and 5G connectivity can be located inside or outside the HMD hardware, as shown in the figure. In the case of AR, the difference between XR5G-A3, XR5G-A4, and XR5G-A5 is the device form factor: as more advanced technology becomes available, we expect to move from a bulky HMD (XR5G-A3) to a small device with minor differences to normal eyeglasses (XR5G-A5). More detailed explanations can be found in~\cite{3GPPRef}.
}

There also exists a thorough discussion in~\cite{3GPPRef} on possible XR use cases and how future networks can help realize them.
%\noindent\textbf{Application use cases.} There exist multiple XR use case examples in~\cite{3GPPRef}. 
These use cases lie in seven broad categories, as shown in Table~\ref{tab:use_cases}. 
Our considered scenario, as explained in Section~\ref{section:xr_scenario}, can cover various use cases from different categories. One prominent example is an XR gaming application, where the users wear XR headsets and are physically co-located in the same application environment. We will discuss the use cases that can benefit from CC in more detail in Section~\ref{sec:EnhancedXR}.

%We will revisit this later when we discuss how coded caching techniques can help improve the XR experience.
\begin{table*}[ht]
    \centering
    \begin{tabular}{m{0.22\textwidth}|m{0.73\textwidth}}
         \textbf{Use case category} & \textbf{Use case example} \\
         \hline
         \hline
         Offline sharing of 3D objects & Alice downloads a 3D model of a sofa from an online shop. She puts the virtual representation of the sofa in her house and checks if it fits nicely. \\
         \hline
         Real-time XR sharing & Alice downloads a 3D model of a sofa and puts the virtual representation of the sofa in her house. Then, she initiates a live discussion with a friend about that. \\
         \hline
         XR multimedia streaming & Bob is watching a live sports game while wearing an XR headset. He can choose where he sits in the stadium and interact with the people sitting nearby. \\
         \hline
         Online XR gaming & Alice and her friends engage in a multiplayer XR game. Each of them can be physically present in a dedicated application hall or just engage remotely. \\
         \hline
         XR mission critical & Police has an important mission. Officers engaging in the mission get live data overlaid on their AR glasses. The data depends on their location and helps with the mission. \\
         \hline
         XR conferencing & A scientific conference is held in a hybrid physical/virtual mode. All the participants wear XR headsets and have the perception of really being at the same place.\\
         \hline
         Spatial audio multiparty call & Bob initiates a group call with his friends. Each participant can share his/her spatial audio and others hear it as if they were physically there. \\
    \end{tabular}
    \caption{XR use case categories and their examples, as provided in~\cite{3GPPRef}.}
    \label{tab:use_cases}
    %\vspace{-10pt}
\end{table*}

%\vspace{-5pt}
\subsection{XR quality of service (QoS) requirements}
%As discussed in~\cite{3GPPRef}, XR applications are expected to put stringent requirements on the networking QoS parameters. This is because they 
XR applications should provide the user with the feelings of \textit{immersion} and \textit{presence}, defined as the feelings of being surrounded by, and physically and spatially located in the virtual environment, respectively. Providing such feelings necessitates very fast and accurate position and orientation tracking and very high-data-rate and low-latency communications. Putting in numbers, an acceptable XR experience requires sub-centimeter positioning accuracy, quarter-degree rotation tracking accuracy, 8K per-eye video resolution, and motion-to-photon latency in the order of 50 milliseconds (the exact number varies by the application category)~\cite{3GPPRef}. 
%Meeting these QoS constraints for a wireless XR application requires a dependable communication infrastructure and careful process- and render-split architectures design. 
Of course, the 5G standard already includes a QoS provision model using the 5G QoS identifier (5QI) parameter. However, many open questions still exist on how the general and graphical processing operations should be split among the client devices and edge servers and which 5QI parameters should be used for each application category. 
%Also, specific standardization support is required for handling the envisioned XR scenario where users are spatially co-located and simultaneously request large amounts of delay-sensitive data.

%stringent QoS parameters, even the sophisticated 5G infrastructure would be challenged to meet the demands. As we discuss later, coded caching techniques can help solve this issue by providing a unique opportunity to benefit from the correlation among the data requested by various users.

%\vspace{-5pt}
\subsection{Rendering and process split}
\label{section:3GPP_render_split}
%\todo[inline]{This section should be changed to reflect how the proposed CC scheme imposes rendering constraints: in order to reduce the delay and improve the QoE, we want to use CC, but it requires some local rendering at the device. Still, the energy expenditure is expected to be [much] smaller than full rendering at UE...}
%\todo[inline]{In the new text, 'energy expenditure' mostly refers to the energy used for rendering, and not the energy required for wireless transmissions. This should be fixed later.}
%In computer graphics, rendering means the process of generating a photo-realistic 3D image from a model. 
In the context of XR, we use the term rendering to refer to the process of generating animated 3D graphics from abstract computer-language models. The software responsible for rendering is called the rendering engine and is part of the higher-level XR engine, which is the software development environment used for building XR applications. The two best-known rendering techniques are rasterization and ray-tracing, whose technical explanation is out of scope for this paper. Ray-tracing generates more realistic outputs but is also more computationally intensive. The selection of the rendering technique and various parameters that can be tuned for the selected technique affects the type and number of data buffers to be processed and the order in which they should be processed. As a result, it also affects the possibility of splitting the rendering process among the local XR device and the edge infrastructure, which is vital for XR applications as XR devices generally have limited processing capability due to limitations in processing power, battery capacity, and allowed power dissipation level.

%In addition to the rendering engine, the way 3D image and video data is encoded and transmitted also affects several critical design parameters. While advanced 2D codecs have been around for a long time, 3D codecs have received less attention due to a lack of applicability. Now, XR applications are opening new perspectives to why 3D contents should be captured, transported, and generated efficiently. Of course, a quick option that is already in use is to mathematically map 3D images to the 2D space and use the existing codes. However, parallel work is also currently carried out to design optimized codecs originally for the 3D content~\cite{3GPPRef}. 

An important question in process splitting, affecting both the energy consumption and user QoE, is where to generate the final image of the FoV shown to the user's eyes.
Fully rendering at end devices is practically infeasible due to the large energy expenditure of the underlying processing, and offloading all the rendering to the edge server makes the QoE prone to the smallest variations in communication latency. Solutions to this issue include splitting the rendering process between the edge and the device or rendering mainly at the edge and making final corrections (following the user's instantaneous pose) at the end device~\cite{3GPPRef}.
In this paper, we promote a modified version of the latter solution: we envision using the edge server for rendering an omnidirectional 3D representation of the static part of the content. The generated 3D images are then transferred, using efficient multi-antenna coded caching techniques, to the end devices which are responsible for overlaying the dynamic and static parts as well as making the final pose corrections locally (see Figure~\ref{fig:static_dynamic_decomposition}).\footnote{As explained earlier, a notable part of the dynamic content can also be delivered with coded caching techniques. Also, overlaying dynamic and static parts can be done, e.g., as explained in~\cite{thomas2020mpeg}.}
Of course, such a process splitting architecture necessitates a higher energy consumption for processing than fully rendering at the edge, as well as larger communication overheads for delivering the 3D content~\cite{3GPPRef}. Nevertheless, the QoE would be more robust to latency variations compared with full rendering at the edge (as minor instantaneous corrections are carried out locally), and the energy expenditure of the underlying processing would be much smaller than full rendering at the device. Moreover, the communication overhead could be efficiently alleviated using novel multi-antenna CC techniques.

\section{Coded Caching for Enhanced XR}\label{sec:EnhancedXR}
Ideally, multi-antenna CC has great potential to provide significant performance improvements for next-generation XR applications. However, critical practical impediments should be resolved before it can be deployed in practice and/or considered for standardization. The most prominent issue is the lack of a proper framework defining the content structure and the timeline for placement/delivery phase operations. In this section, we first review a few application use cases that lie within our envisioned system model in \rev{Section~\ref{section:intro}.\ref{section:xr_scenario}}. Then, we propose a possible framework for integrating CC techniques in such use cases and review a few issues open for future study.

\begin{table*}[ht]
    \centering
    \small{
    \begin{tabular}{c||c|c|c|c|c}
         \textbf{Parameter} & \textbf{Scenario I} & \textbf{Scenario II} & \textbf{Scenario III} & \textbf{Scenario IV} & \textbf{Scenario V} \\ 
         \hline
         \hline
         Application environment size & $5m \times 5m$ & $5m \times 5m$ & $5m \times 5m$ & $10m \times 10m$ & $10m \times 10m$ \\
         \hline
         User count & 5 & 10 & 10 & 10 & 40 \\
         %\hline
         %STU size for 3D reconstruction & $50cm \times 50cm$ & $50cm \times 50cm$ & $50cm \times 50cm$ & $50cm \times 50cm$ & $50cm \times 50cm$ \\
         \hline
         Library file count & 100 & 100 & 100 & 400 & 400 \\
         %\hline
         %3D tile file size & 100 MB & 100 MB & 100 MB & 100 MB & 100 MB \\
         \hline
         User cache size & 4 GB & 4 GB & 8 GB & 8 GB & 8 GB \\
         \hline
         \rev{The coded caching gain ($t$)} & \rev{2} & \rev{4} & \rev{8} & \rev{2} & \rev{8} \\
         \hline
         \rev{CC packet size} & \rev{5 MB} & \rev{10 MB} & \rev{20 MB} & \rev{20 MB} & \rev{$\sim$ 20 KB} \\
         \hline
         Transmitter \rev{spatial multiplexing gain $(L)$} & 2 & 2 & 2 & 2 & 2\\
         \hline
         Parallel streams w/ CC \rev{($t+L$)} & 4 & 6 & 10 & 4 & 10\\
         \hline
         \textbf{Improvement by CC} & \textbf{100\%} &\textbf{ 200\%} & \textbf{400\%} & \textbf{100\%} & \textbf{400\%} \\
    \end{tabular}
    }
    %\vspace{-2mm}
    \caption{Theoretical CC gains in different XR setups, $50cm \times 50cm$ STU size, 100 MB of 3D image size}
    \label{tab:my_label}
\end{table*}

%\vspace{-5pt}
\subsection{CC and application use cases}
CC suits well to XR applications where the users have physical proximity, can move freely, and require location-dependent data. Moreover, the requested content (or part of it) should be cacheable by nature (as thoroughly discussed in Section~\ref{section:xr_scenario}).
%and separated from the dynamic part (as shown in Figure~\ref{fig:static_dynamic_decomposition}). 
%For example, in many interactive applications, a major part of the requested data, used for reconstructing the 3D model (infrastructure) of the virtual environment, does not change frequently in time and can be cached proactively. In this paper, we assume appropriate software models are used to separate these static and dynamic parts of the requested data.
%, and the interactive nature of the application necessitates ultra-low communication latency.
Some examples of XR applications where CC can be applied are:
%Wireless XR applications are growing in popularity. They will put stringent requirements on networking parameters such as data rate and delay, and in specific scenarios, meeting these requirements can be beyond the capability of currently state-of-the-art networks. This is especially true for group XR activities where a number of freely moving users engage in the same XR application. Such activities lie into the envisioned model described in Section~\ref{section:xr_scenario} and include interesting use cases such as:
%\begin{itemize}

\noindent\ding{227} \textbf{An XR gaming application} where the users are scattered in the game hall, equipped with XR headsets, and move freely. The users play an interactive game, and their requested data at any moment depends on their location at that moment;

\noindent\ding{227} \textbf{An XR museum application} where the users entering the museum are given XR headsets and move freely throughout the museum building. Close to each historical item, nearby users may jointly enter a related virtual 3D world where they can interact with virtual elements;

\noindent\ding{227} \textbf{An XR conferencing application} where the users may join physically or virtually. In both cases, the users are equipped with XR headsets and move freely among different halls. The users can interact with each other and with the environment.
%\end{itemize}

All these use cases share common properties; the users have physical proximity, they can move freely and require location-dependent data, and the interactive nature of the application necessitates ultra-low communication latency. However, even with recent advancements, it is challenging to simultaneously transmit very high data rate 3D streams with ultra-low latency to multiple users. The situation becomes even more challenging as the users start to move, as the stringent \rev{requirements on rate and latency} should be met throughout the whole application environment.

%Here we detail how coded caching techniques help alleviate these challenges by efficient multicasting of (part of) the multimedia content to multiple users. It should be noted that, similar to any other caching method, coded caching can only be applied to the static part of the data that does not frequently change with time and is common for multiple users. However, in all the uses case mentioned above, a major part of the data content is needed to reconstruct the 3D model (infrastructure) of the virtual environment. In order to use coded caching techniques for this static content, appropriate software development models should be used to separate it from the frequently-changing dynamic part of the data. In this paper, we assume the separation between the static and dynamic parts is in place.

\rev{

\subsection{Render and process split for CC-aided XR}
In Section~\ref{section:3GPP}.\ref{section:3GPP_render_split},
after briefly reviewing render and process split frameworks considered by 3GPP, we
%we briefly discussed render and process split models considered by 3GPP. We also
promoted a model where rendering the omnidirectional 3D representation of the static part of the content is done at the edge server, and overlaying the dynamic part and final pose corrections are done locally at the end device. This model makes the QoE less prone to small variations in transmission delay but requires delivering larger amounts of data, which could be alleviated using multi-antenna CC techniques.

We must emphasize that the promoted render and process split model is \emph{not} the only possible way to use CC techniques to enhance QoE in XR applications. In fact, we may always use CC techniques when (part of) the delivered content is cacheable in nature. Indeed, there exist multiple works in the literature for efficient rendering and delivery of omnidirectional image and video for XR use cases, with a comprehensive survey provided in~\cite{Yaqoob2020AOpportunities}. For example, considering the tile-based rendering model where an omnidirectional frame is split into multiple tiles and only tiles relevant to the FoV of the user are delivered~\cite{Yaqoob2020AOpportunities}, the static background of the image in each tile is again cacheable and could be efficiently delivered with CC techniques. Even if we intend to deliver the tiles closer to the center of the FoV with a higher quality to save bandwidth (as those tiles have a more prominent effect on the QoE of the user), we may use novel CC techniques, e.g., based on multiple descriptor codes (MDC)~\cite{goyal2001multiple,salehi2020coded}, to support different image qualities for various tiles.

Similarly, as long as (part of) the transmitted data is cacheable in nature (or could be split into cacheable elements as discussed for the dynamic part of the content in Section~\ref{section:intro}.\ref{section:xr_scenario}), CC schemes provide benefits even if disruptive techniques such as semantic communication are used to deliver the content~\cite{Lan2021WhatIntelligence,Chaccour2022LessNetworks}, or if multi-sensory XR techniques are applied to further improve users' immersion into the virtual world. For example, with semantic communications, the cacheable part could include the semantic language database.

%One may argue the effectiveness of CC techniques if disruptive techniques such as semantic communications~[XXX] are used to deliver the content requested by users in an XR setup. Of course, such techniques are in the early phases of development and their true realization is yet to arrive. Nevertheless, CC techniques could still be used to efficiently deliver every part of data that is cacheable in nature. This can include, e.g., the semantic language data model used for communication.
}

%\vspace{-5pt}
\subsection{The envisioned XR framework}
\label{section:zoning_stu}

\begin{figure}[t]
\begin{center}
    \includegraphics[width=0.9\columnwidth]{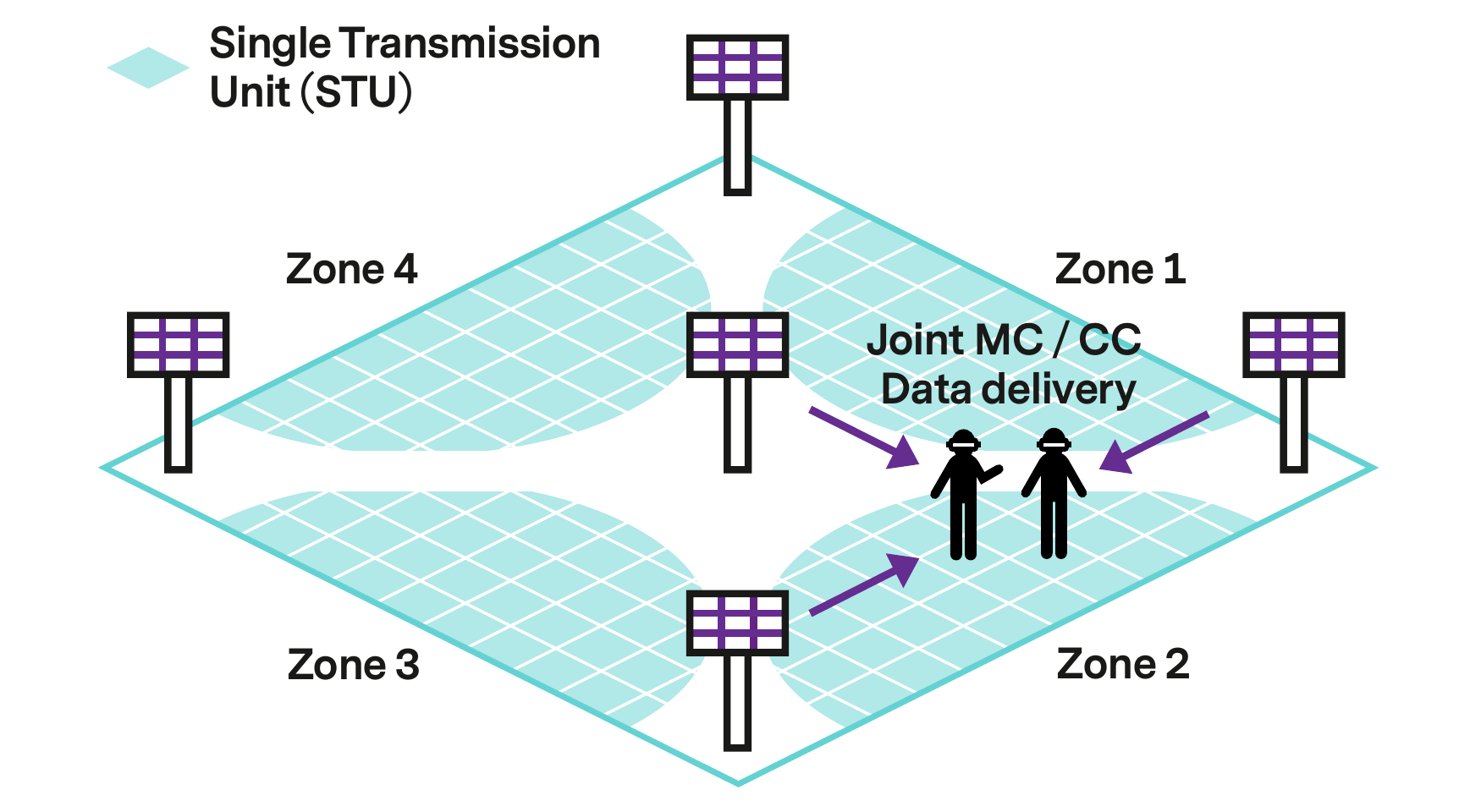}
    \caption{Zoning and STUs for improved XR}
    \label{fig:zoning}
    %\vspace{-15pt}
\end{center}
\end{figure}

As discussed earlier, for the considered collaborative XR setup, we propose 
%separating identifying cacheable parts of the content
%static and dynamic parts of the content 
and using multi-antenna CC techniques to deliver them to the end users efficiently.
As discussed, these cacheable parts include, for example, the static part of the FoV as well as the building elements of the dynamic part. To keep discussions simple, here we assume CC-aided data delivery is applied only to the static part.
%Moreover, t
To reduce the impact of transmission delay variations on QoE, we assume omnidirectional 3D images are used to represent the static part of the content, and final pose corrections are done at the end device. In this regard, we define the \textbf{single transmission unit (STU)} to be the smallest area unit for which the static view can be transmitted with a single 3D image. For example, we can assume the XR application environment is split into tiles of size 50$\times$50 cm and each tile is an STU, i.e., the user needs to receive a new 3D image from the server to reconstruct the background scenery as it moves from one tile to another (see Figure~\ref{fig:zoning}).

Delivering high-definition omnidirectional 3D images in a collaborative XR application necessitates transmitting large amounts of data (hundreds of megabytes) in a very short time frame (tens of milliseconds). Nevertheless, this burden could be handled through the efficient usage of novel multi-antenna CC techniques. Table~\ref{tab:my_label} represents theoretical gains (i.e., upper bounds on the achievable performance) in terms of the number of possible interference-free parallel data streams resulting from CC techniques in a few example XR setups.
This table highlights why CC could be an exciting solution for future XR setups: it enables a large performance gain under realistic assumptions, and this gain scales with the number of users. This is in contrast to many other communication techniques, where the performance primarily deteriorates as the network size scales.\footnote{\rev{
It should be noted that the numbers in Table~\ref{tab:my_label} are indicative; used only to provide a general idea of the gains possible by CC techniques. 
%Moreover, the term `transmitter multiplexing gain' refers to the number of parallel streams that can be supported by the multi-antenna transmitter, which is less than or equal to the number of antennas at the transmitter. Indeed, with the recent emergence of communications in higher frequencies (e.g., in mmWave bands), it has become possible to employ larger antenna arrays ($>100$) at both transmitters and receivers. However, the attainable multiplexing gain is still limited in practice due to limitations in the number of RF chains and the available baseband processing power, and this is the main reason for choosing a small multiplexing gain in Table~\ref{tab:my_label}. Finally, 
Also, to calculate the CC packet size, we have considered the CC scheme of~\cite{lampiris2018adding} for its reduced subpacketization (except for Scenario~I, where the CC scheme of~\cite{shariatpanahi2018physical} is used as the scheme of~\cite{lampiris2018adding} incurs a performance loss).
}} 

One might argue the feasibility of the proposed framework in very large XR environments (e.g., within a large exhibition or conference hall) as, in such a case, users could be far from each other and served by different transmitters. For this case, we further propose splitting the whole environment into several \textbf{zones}, where the users within the same zone could participate in a CC-aided content delivery session and the cache contents of users are updated as they move between the zones (see Figure~\ref{fig:zoning}). Of course, one should note that although zoning allows allocating larger cache portions to each cached file (as it limits the number of files by constraining the physical area), it does not necessarily increase the achievable CC gain. This is because the CC gain is proportional to the total cache size of the users in the same CC-aided delivery session~\cite{maddah2014fundamental}, and zoning limits the expected number of such users.

Another important issue is the overall \textit{energy efficiency} of the proposed framework.
\rev{Of course, studying the energy efficiency of the CC schemes is not limited to XR setups and is addressed in a limited number of works in the literature. One notable example is~\cite{Vu2018Edge-cachingOptimization}, where the energy efficiency of the delivery phase is compared with and without using CC schemes. However, for our proposed scheme, studying energy efficiency should also consider the placement phase, as
%It is caused by the proactive placement phase, where 
the users have to download large amounts of data with corresponding energy expenditure to fill up their cache memories, while part of the downloaded content might never be used in the delivery phase.} The zoning further increases the placement cost, as the users might need to update their cache contents frequently while moving between the zones.
%Naturally, the energy efficiency problem of CC schemes is not limited to XR setups, and to the best of our knowledge, a thorough study of its severity and possible workarounds is still lacking in the literature. Nevertheless, for our considered XR scenario, 
One
%One 
way to address this issue is to optimize the zoning process
such that the users pass through the vicinity of TRPs as they move from one zone to another; so that they can promptly download the data required for updating their cache contents at high rates and hence causing limited overall energy expenditure.
%in order to maximize the chance of users being close to TRPs (hence, being able to receive data more efficiently) when their cache contents are updated. 
Moreover, it is also possible to optimize the XR application (using the available information on how the zoning is performed) to increase the expected time each user resides within a single zone, thus decreasing the need to update the cache contents and increasing the number of time cache contents are reused. 
\rev{As a clarifying example, consider an XR application where two groups of players are engaged in a first-person action game. Then, noting that the users spend more time in designated conflict areas in such an application, the game map could be altered such that each conflict area lies within a single CC zone.
}

\rev{As a final note, although most CC schemes in the literature assume synchronized user requests (which is not a strong assumption for XR use cases), it is also possible to apply CC schemes without such a constraint. For example, one may use the CC scheme in~\cite{lampiris2021coded}, for which the achievable performance is within a multiplicative factor of two from the case of the synchronized request under the assumption of uncoded cache placement.}

\begin{figure}[t]
    \centering
    \resizebox{0.75\columnwidth}{!}{%
    
    \begin{tikzpicture}

    \begin{axis}
    [
    % put axis lines at left and bottom
    axis lines = left,
    % control axis labels
    xlabel = \smaller {Total transmission time (ms)},
    xtick={4,6,8,10,12,14,16,18,20},
    xmin = 4,
    xmax=22,
    xmode=log,
    log ticks with fixed point,
    x filter/.code=\pgfmathparse{#1},
    ylabel = \smaller {CDF},
    ylabel near ticks,
    % control legend position
    legend pos = south east,
    % control size of tick marks (10,20,30,etc)
    ticklabel style={font=\smaller},
    % control major grids
    grid=major,
    %major grid style={line width=.2pt,draw=gray!30},
    % control minor grids
    %grid style={line width=.1pt, draw=gray!10},
    %minor tick num=3,
    ]
    
    % \addplot[black]
    % table[y=Opt-L1-Y,x=Opt-L1-X]{Figs/CDF_data_L.tex};
    % \addlegendentry{\smaller $L=1$, Analytical}
    % \addplot[black,dashed]
    % table[y=Super-L1-Y,x=Super-L1-X]{Figs/CDF_data_L.tex};
    % \addlegendentry{\smaller $L=1$, Super-user}
    % \addplot[black!50]
    % table[y=Heur-L1-Y,x=Heur-L1-X]{Figs/CDF_data_L.tex};
    % \addlegendentry{\smaller $L=1$, RGA}
    
    % \addplot[green]
    % table[y=Opt-L5-Y,x=Opt-L5-X]{Figs/CDF_data_L.tex};
    % \addlegendentry{\smaller $L=5$, Analytical}
    % \addplot[green,dashed]
    % table[y=Super-L5-Y,x=Super-L5-X]{Figs/CDF_data_L.tex};
    % \addlegendentry{\smaller $L=5$, Super-user}
    % \addplot[green!50]
    % table[y=Heur-L5-Y,x=Heur-L5-X]{Figs/CDF_data_L.tex};
    % \addlegendentry{\smaller $L=5$, RGA}

    \addplot[gray,dashed]
    table[y=Unicasting_Uniform_CDF,x=Unicasting_Uniform]{Figures/Location_dependent_CC_data.tex};
    \addlegendentry{\smaller Uniform Unicast}
    \addplot[gray]
    table[y=Single_User_Placement_NoCC_CDF,x=Single_User_Placement_NoCC]{Figures/Location_dependent_CC_data.tex};
    \addlegendentry{\smaller Non-uniform Unicast}
    %\addplot[blue]
    %table[y=Multi_User_Placement_NoCC_CDF,x=Multi_User_Placement_NoCC]{Figures/Location_dependent_CC_data.tex};
    %\addlegendentry{\smaller No-CC Multi User}
    \addplot[black,dashed]
    table[y=Uniform_Placement_CC_CDF,x=Uniform_Placement_CC]{Figures/Location_dependent_CC_data.tex};
    \addlegendentry{\smaller Baseline CC}
    \addplot[black]
    table[y=Proposed_CDF,x=Proposed]{Figures/Location_dependent_CC_data.tex};
    \addlegendentry{\smaller Non-uniform CC}
    %\addplot[gray]
    %table[y=Opt-L20-Y,x=Opt-L20-X]{Figs/CDF_data_L.tex};
    %\addlegendentry{\smaller $L=20$}

    \end{axis}

    \end{tikzpicture}
    }
    \caption{The CDF of total delivery time (logarithmic-scale) for $K = 36$ users, $\gamma = 0.33$, and $L = 6$. The variance of the shadowing effect of the channel is $\sigma_s = 7$~\cite{mahmoodi2023multi}.}
    \label{fig:cdf_location_dependent}
\end{figure}
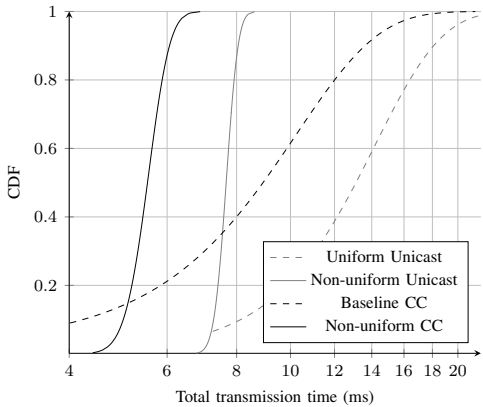

\rev{
\subsection{Location-dependent coded caching for XR}
\label{section:location_dependent}
%\noindent\textbf{Location-dependent cache placement.}
%The performance of the proposed cache-aided communication strategy can be further improved through a non-uniform cache placement strategy.

As discussed in Section~\ref{section:intro}, XR applications possess two important features: location-dependent content requests and file libraries of limited size.
%: the content requested by a user at a time moment depends on the user's physical location (i.e., the STU index) in that time moment. On the other hand, as the general structure of the application hall is fixed, the large-scale fading of the wireless channel is also predictable at every STU location. As a result, we face a location-dependent content request scenario with an estimation of the channel strength when a content file is requested. 
These features are used in~\cite{mahmoodi2021non,mahmoodi2022asymmetric,mahmoodi2022non,mahmoodi2023multi} to design new CC schemes well-tailored to XR setups. The core idea is to allocate larger cache portions for storing (parts of) the content requested in STUs with poor channel connectivity to avoid excessive content delivery delays and improve QoE. Then, novel CC techniques are introduced to provide a global caching gain for the resulting non-uniform memory allocation. 
Of course, multi-user content delivery in such schemes requires delivering different-sized data chunks to various users within a single transmission, which is done using either nested code modulation (NCM)~\cite{tang2017coded} or a high-performance beamformer design~\cite{mahmoodi2023multi}. 
Here, to showcase the QoE improvements brought to XR applications by CC techniques, in Figure~\ref{fig:cdf_location_dependent}, we have provided simulation results from~\cite{mahmoodi2023multi} to compare the following four schemes: 1) the location-dependent CC scheme of~\cite{mahmoodi2023multi}, denoted by `Non-uniform CC,' 2) the baseline multi-antenna scheme of~\cite{shariatpanahi2018physical}, denoted by `Baseline CC,' 3) a reference scheme with non-uniform memory allocation but unicast content delivery (i.e., no CC technique used), denoted by `Non-uniform Unicast,' and 4) another reference scheme with uniform memory allocation and unicast content delivery, denoted by `Uniform Unicast.' This figure clarifies the general impact of coed caching techniques as well as the QoE benefits of location-dependent CC schemes: the variance in the content delivery time is much smaller than the baseline CC scheme due to the underlying non-uniform memory allocation, and the achievable rate is better than the unicast case due to the global caching gain of CC techniques.

Of course, we should note that other works also exist in the literature on location-dependent coded caching~\cite{wan2022optimal}. However, in this paper, we have limited our review to~\cite{mahmoodi2021non,mahmoodi2022asymmetric,mahmoodi2022non,mahmoodi2023multi} as they also consider XR applications as the use case. 

%The results of these works, replicated partly in Figures~XX and~XX here, clearly emphasize the QoE benefits of CC schemes in XR applications. \textbf{A FEW SENTENCES ON THE RESULTS.} It should be noted that the above-mentioned works are not the only ones discussing CC gains in applications with location-dependent content requests~[XX]. However, in this paper, we limit our focus to XR-related results.

%By allocating larger cache portions to contents associated with STUs with poor channel conditions, users in poor-connectivity locations would require less amount of data from the edge server and would enjoy an enhanced QoE compared to the uniform cache placement scenario.
%Two notable works in this direction are recently proposed in~\cite{mahmoodi2021non,mahmoodi2022non}, for single- and multi-antenna setups. In both works, cache allocation for each content file is assumed to be proportional to the achievable rate at its respective location in the application environment, and novel delivery algorithms are used to enable an additional CC gain. 

%This additional caching gain is achieved by combining the variable-size data streams requested by different users within a single multicast message, using nested code modulation (NCM) techniques. Extensions to more practical multi-transmitter and multi-antenna setups are needed before this idea can be used in practice.
}

\begin{figure}[t]
\centering
\resizebox{0.75\columnwidth}{!}{%

\begin{tikzpicture}

\begin{axis}
[
axis lines = left,
xmin=0,
%xlabel near ticks,
xlabel={\smaller Standard deviation of the user-profile assignment ($\sigma$)},
ymin=0.59,
ymax=1.02,
ylabel={\smaller DoF},
%ylabel near ticks,
    % control major grids
    grid=both,
    major grid style={line width=.2pt,draw=gray!30},
    % control minor grids
    grid style={line width=.1pt, draw=gray!10},
    minor tick num=5,
    legend pos = south east,
ticklabel style={font=\smaller},
]

\addplot[only marks, mark=x, mark size=2.2pt, black!50] table[y=DynamicY,x=DynamicXSD]{Figures/Dynamic_Data_2.tex};
\addlegendentry{\smaller Dynamic CC}

\addplot [gray,dashed,line width=1.0pt]
table[y=UCY,x=UCXSD]{Figures/Dynamic_Data_1.tex};
\addlegendentry{\smaller Unicast Only}

\addplot [black,line width=1.0pt]
table[y=MCY,x=MCXSD]{Figures/Dynamic_Data_1.tex};
\addlegendentry{\smaller Uniform CC}

\end{axis}

\end{tikzpicture}
}
\caption{The average number of parallel streams per transmission (DoF) vs non-uniformity in the user-profile assignment. $K=50$ users, $\gamma = 0.1$, $L=9$, ten cache profiles.}
\label{fig:Dynamic}

\end{figure}
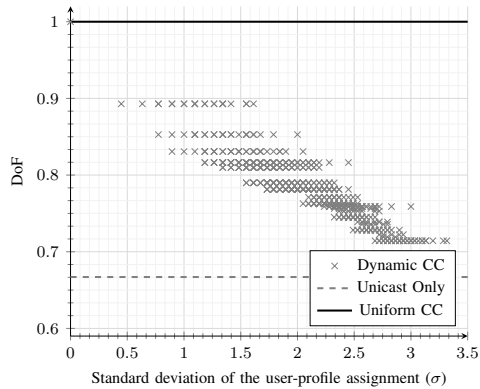

\rev{
\subsection{Coded caching for dynamic setups}
\label{section:dynamic_cc}

From the cache placement design perspective, CC schemes range from fully centralized schemes, where a central server instructs what should be cached by every individual user, to fully decentralized ones, where the users randomly cache arbitrary portions of the content files. 
Fully decentralized schemes have the important advantage of being flexible to variations in network parameters
%Decentralized schemes are more flexible in handling uncertainties in network parameters 
but provide comparable performance to fully centralized schemes only asymptotically (i.e., when the number of users is very large)~\cite{maddah2015decentralized}. 
This has led to the design of \textit{shared-cache} CC models, which are centralized in the sense that the cache contents of the users are determined by a set of predefined \emph{cache profiles} but decentralized in the sense that the assignment of the users to cache profiles could be random (it is even possible that multiple users are assigned to the same profile and have similar cache contents)~\cite{parrinello2019fundamental,parrinello2020extending,jin2019new,Parrinello2023FundamentalNetworks,Dutta2021DecentralizedCaches}. 
All shared-cache CC models are signal-level CC schemes and have similar strong (e.g., reduced subpacketization) and weak (e.g., inferior finite-SNR performance) points.
Due to their great flexibility, shared-cache models have also been used to build CC schemes for dynamic networks where the users may join and leave the network freely, with thorough analyses available in the literature for both high-SNR~\cite{abolpour2022coded,abolpour2023cache} and finite-SNR~\cite{salehi2021low} regimes. In Figure~\ref{fig:Dynamic}, to showcase the performance of dynamic CC schemes, we have provided simulation results from~\cite{abolpour2022coded} that compare the performance of their proposed scheme with the two baseline schemes of unicasting only (no CC scheme applied, denoted by Unicast Only) and no dynamic conditions (CC scheme of~\cite{salehi2021low} applied, denoted by Uniform CC). As can be seen, the performance of the dynamic CC scheme lies within the two baselines and gets better as users are assigned to the cache profiles more uniformly.

In the context of CC schemes for wireless XR applications, even if the number of users is assumed to be fixed and known during the application runtime, we still face a dynamic setup for two good reasons: 1) with the zoning definition, CC data delivery is performed within each zone separately, and the number of users within a zone is not fixed as the users move between the zones, and 2) it is possible that a subset of users do \emph{not} initiate any data requests in a time interval, e.g., if they have not changed their location from the previous interval. As a result, it is worthwhile to design dynamic CC schemes specially tailored to XR setups.
}

\subsection{Further performance improvements}

\noindent\textbf{D2D-assisted coded caching.}
The CC gain is a result of cache-aided interference cancellation at the receivers.
%In coded caching, carefully created codewords are transmitted to different subsets of users. Part of the interference is avoided over-the-air with beamforming techniques, and the remaining part is removed using the cache contents of the receiving user. 
However, from another viewpoint, this interference removal can be regarded as \textit{exchanging} data between multiple users. For example, in the single-antenna CC example in the left-hand side of Figure~\ref{fig:cc_illustration}, it can be imagined that the transmitted codeword $A_2 \oplus B_1$ has enabled user one to deliver $B_1$ (cached in its memory) to user two, and at the same time, user two to deliver $A_2$ to user one. Now, instead of using multicasting to deliver the codeword $A_2 \oplus B_1$, two nearby users can exchange the requested terms using a direct device-to-device (D2D) link with higher capacity and lower latency~\cite{ji2015fundamental,mahmoodi2020d2d,mahmoodi2023d2d}. %Investigating possible performance improvements of D2D-assisted CC is part of the ongoing research by the community.

\noindent\textbf{Optimizing the zoning process.}
As discussed, zoning is a necessary tool for applying the proposed framework to large XR environments, \rev{and the way we perform zoning affects the energy efficiency of the system.}
%The way we perform zoning affects the energy efficiency of the system, as the users need to download large amounts of data to update their cache contents as they move from zone to zone. 
In this regard, optimizing the zoning process to ensure that the users pass through the vicinity of TRPs while changing zones could improve the energy efficiency of the system noticeably.
\rev{An important question arising here is whether to allow any overlap among the zones. With overlaps, the number of cache update operations might be increased, but users have more time to update their cache contents as they move between the zones, enabling more efficient resource allocation for the content update process and reducing the probability of incomplete cache updates. Studying the trade-off between the performance and energy efficiency of the network as the size of the overlap areas is increased is out of the scope of this paper. Also, as mentioned in Section~\ref{sec:EnhancedXR}.\ref{section:zoning_stu},}
%Moreover, 
for a given zoning, we may also improve energy efficiency by optimizing the XR application such that the users are less likely to change their zones within short time intervals. The result is an improved reuse factor of the cache contents, as the users spend more time and request a larger number of contents within each zone. 
\rev{Indeed, such optimizations of the XR application also affect the above-mentioned trade-off.}

%Nevertheless, such ideas are not yet mature as, to the best of our knowledge, a thorough end-to-end study on the energy efficiency of CC schemes in XR setups is still missing and needs to be addressed in the literature.
%such ideas to improve energy efficiency are not yet mature as, to the best of our knowledge, there exists no comprehensive study in the literature on the energy efficiency of CC setups in XR environments.
%However, the way we perform zoning also determines the energy efficiency of the proposed solution as it affects the number of times the cache contents of the users are updated. Optimizing the zoning process for improving energy efficiency is an open problem and requires additional knowledge, such as users' movement patterns that could be collected through various methods (e.g., using machine learning techniques).

%\todo[inline]{Energy expenditure discussions - optimizing with learning techniques.}

%\vspace{-5pt}
\subsection{Selecting the proper scheme}
\label{section:signal_bit_level_selection}
An important implementation aspect is the choice of a suitable CC scheme for the target XR application scenario, as the schemes vary widely from the performance, complexity, and signaling overhead perspectives. Especially the choice determines the underlying interference cancellation approach. Unlike signal-level schemes, bit-level schemes allow the cache-aided interference cancellation to be carried out at higher network layers, making the process transparent to the physical layer and hence, easier to implement.
%for lower communication layers, i.e., physical layer and the medium access control (MAC). 
%Moreover, multicasting is supported with bit-level approaches using an appropriate beamformer design, and hence, the impact on processing-intensive lower layers can be minimized or even avoided. 
However, as discussed earlier, the bit-level approach is more sensitive to the subpacketization problem, leading to either reduced CC gain or an extreme number of subpackets with increased complexity on signaling and the total number of individual transmissions.

On the other hand, signal-level schemes highly impact the physical later
%
%The signal-level approach benefits from much smaller subpacketization but it has a high impact on physical layer processing
as the interference cancellation is performed before decoding the data, 
%(i.e., in the signal domain). T
requiring the receiver to fetch the data to be canceled from its memory and to regenerate a replica of the expected interference by carrying out channel encoding, rate matching, scrambling, and modulation locally. The receiver also needs to estimate the effective channel, incorporate the effect of the receiver beamformer and equalizer, and convolute the generated signal with the effective channel for all interferers to be canceled. 
%On the other hand, signal-level interference cancellation 
Nevertheless, the signal-level approach can facilitate a user-specific link adaptation in terms of modulation and coding and hence, may provide a way to alleviate the near-far problem hindering the CC operation.

\section{Conclusion}
%Creating and consuming new virtual or digital twin worlds will be one of the key 6G drivers. Numerous XR applications will emerge with use cases in education, industry, gaming, and social networking, and will require stringent performance parameters from the underlying wireless infrastructure. Specifically, in collaborative XR applications, the user experience will be hindered by the sporadic and unreliable wireless connectivity, and hence, new technology components will be needed to provide more evenly distributed QoE.

%In this paper, w
We explored novel multi-antenna coded caching techniques as an appropriate candidate for resolving wireless connectivity challenges of future collaborative XR applications. We first reviewed recent advancements in multi-antenna coded caching techniques and then discussed how XR application requirements are addressed within the 3GPP framework. Finally, we identified new challenges arising from integrating CC techniques into multi-user XR scenarios and proposed novel solutions to address them in practice.

\bibliographystyle{IEEEtran}
\bibliography{references,3GPPRef,ref_whitepaper}

\newpage

%\vspace{-35pt}
\begin{IEEEbiographynophoto}
%[{\includegraphics[width=1in,height=1.25in,clip,keepaspectratio]{Figures/Authors/MJS.jpg}}]
{MohammadJavad Salehi } Received his B.Sc., M.Sc., and Ph.D. in Electrical Engineering from the Sharif University of Technology, Tehran, Iran, in 2010, 2012, and 2018, respectively. Since 2019, he has been
%Currently, he is 
a postdoctoral researcher at the Center for Wireless Communications (CWC), University of Oulu, Finland. His research interests include coded caching and multi-antenna communications.

\end{IEEEbiographynophoto}

%\vspace{-35pt}
\begin{IEEEbiographynophoto}{Kari Hooli}
received the D.Sc. (Tech.) degree in Electrical Engineering from the University of Oulu, Finland, in 2003. Currently, he is a Distinguished Member of the Technical Staff at Nokia Standards in Oulu. Before joining Nokia, he worked at the Centre for Wireless Communications (CWC) at the University of Oulu and visited the Centre for Communication Systems Research at the University of Surrey, UK. He holds numerous patents on 4G and 5G technologies and has authored several conference papers, journals, and book chapters. His research interests include physical layer design and signal processing for wireless communications and cellular networks.

\end{IEEEbiographynophoto}

%\vspace{-35pt}
\begin{IEEEbiographynophoto}{Jari Hulkkonen}
is the Department Head, Radio Research at Nokia Standards, Oulu. He graduated in 1999 (M.Sc.EE) from the University of Oulu, Finland. 
He has been working in Nokia since 1996. He started his career at Nokia in GSM/EDGE research and standardization projects. Since 2006 he has been leading radio research in Nokia Oulu. 
Currently, Jari is Radio Research Department Head in Nokia Standards with a focus on the 5G New Radio evolution. He has more than 30 patents/patent applications in 2G-5G technologies as well as several publications and book chapters.
\end{IEEEbiographynophoto}

%\vspace{-35pt}
\begin{IEEEbiographynophoto}{Antti T\"olli}
%(M'08, SM'14)  
is a Professor with the Centre for Wireless Communications (CWC), University of Oulu, Finland. He received the D.Sc. (Tech.) degree in electrical engineering from the University of Oulu in 2008. From 1998 to 2003, he worked at Nokia Networks as a Research Engineer and Project Manager both in Finland and Spain. 
In May 2014, he was granted a five-year (2014-2019) Academy Research Fellow post by the Academy of Finland. 
He has authored numerous papers in peer-reviewed international journals and conferences and several patents. His research interests include radio resource management and transceiver design for broadband wireless communications, with a special emphasis on distributed interference management in heterogeneous wireless networks.
\end{IEEEbiographynophoto}

\end{document}